\documentclass[superscriptaddress,secnumarabic,amssymb,amsmath,nobibnotes,aps,prd,showkeys,nofootinbib,onecolumn,a4paper,12pt]{revtex4}
\usepackage{bigints}
\usepackage{float} 
\usepackage{paralist}
\usepackage{multirow}
\usepackage{array}
\newcolumntype{P}[1]{>{\centering\arraybackslash}p{#1}}
\usepackage[T1]{fontenc}    % use 8-bit T1 fonts
\usepackage{hyperref}       % hyperlinks
\usepackage{url}            % simple URL typesetting
\usepackage{booktabs}       % professional-quality tables
\usepackage{amsfonts}       % blackboard math symbols
\usepackage{nicefrac}       % compact symbols for 1/2, etc.
\usepackage{microtype}      % microtypography
\usepackage{lipsum}		% Can be removed after putting your text content
\usepackage{graphicx}
\usepackage{natbib}
\usepackage{doi}
\usepackage{graphicx}
\usepackage{subfigure}
\usepackage{epsf}
\usepackage{bm}
\usepackage{amsmath}
\usepackage{amsfonts}
\usepackage{amssymb}
\usepackage{graphicx}
\usepackage{tabularx}
\usepackage{color}%
\providecommand{\U}[1]{\protect\rule{.1in}{.1in} }

\newcommand{\be}{\begin{equation}}
\newcommand{\ee}{\end{equation}}

\newcommand{\mincir}{\raise
-3.truept\hbox{\rlap{\hbox{$\sim$}}\raise4.truept\hbox{$<$}\ }}
\newcommand{\magcir}{\raise
-3.truept\hbox{\rlap{\hbox{$\sim$}}\raise4.truept\hbox{$>$}\ }}

\providecommand{\U}[1]{\protect\rule{.1in}{.1in}}

\usepackage{tikz,xcolor,hyperref}

\definecolor{lime}{HTML}{A6CE39}
\DeclareRobustCommand{\orcidicon}{%
	\begin{tikzpicture}
	\draw[lime, fill=lime] (0,0) 
	circle [radius=0.16] 
	node[white] {{\fontfamily{qag}\selectfont \tiny ID}};
	\draw[white, fill=white] (-0.0625,0.095) 
	circle [radius=0.007];
	\end{tikzpicture}
	\hspace{-2mm}
}

\foreach \x in {A, ..., Z}{%
	\expandafter\xdef\csname orcid\x\endcsname{\noexpand\href{https//orcid.org/\csname orcidauthor\x\endcsname}{\noexpand\orcidicon}}
}

\begin{document}

\title{Exact solutions and cosmological constraints in fractional cosmology}

\newcommand{\orcidauthorA}{0000-0001-6769-5722}
\newcommand{\orcidauthorB}{0000-0002-1152-6548}
\newcommand{\orcidauthorC}{0000-0002-9695-3409}

\author{Esteban González \orcidA{}}
\email{esteban.gonzalez@uac.cl}
\affiliation{Direcci\'on de Investigaci\'on y Postgrado, Universidad de Aconcagua, Pedro de Villagra 2265, Vitacura, 7630367 Santiago, Chile}

\author{Genly Leon *\orcidB{}}
\email{genly.leon@ucn.cl}
\affiliation{Departamento de Matem\'{a}ticas, Universidad Cat\'{o}lica del Norte, Avda.
Angamos 0610, Casilla 1280, Antofagasta, 1270709, Chile}
\affiliation{Institute of Systems Science, Durban University of Technology, P.O. Box 1334,
\mbox{Durban 4000, South Africa}}

\author{Guillermo Fernandez-Anaya \orcidC{}}
\email{guillermo.fernandez@ibero.mx}
\affiliation{ Depto. de F\'isica y Matem\'aticas, Universidad Iberoamericana, Ciudad de M\'exico, Prolongaci\'on Paseo  de la Reforma 880, M\'exico D. F. 01219, M\'exico}

\begin{abstract}
This paper investigates exact solutions of cosmological interest in fractional cosmology. Given $\mu$,  the order of Caputo's fractional derivative, and $w$, the matter equation of state, we present specific exact power-law solutions. We discuss the exact general solution of the Riccati Equation, where the solution for the scale factor is a combination of power laws. Using cosmological data, we estimate the free parameters. An analysis of  type Ia supernovae (SNe Ia) data and the observational Hubble parameter data (OHD), also known as cosmic chronometers, and a joint analysis with data from SNe Ia + OHD leads to best-fit values for the free parameters calculated at $1\sigma$, $2\sigma$ and $3\sigma$ confidence levels (CLs). On the other hand, these best-fit values are used to calculate the age of the Universe, the current deceleration parameter (both at $3\sigma$ CL) and the current matter density parameter at $1\sigma$ CL. Finding a Universe roughly twice as old as the one of $\Lambda$CDM is a distinction of fractional cosmology. Focusing our analysis on these results, we can conclude that the region in which $\mu>2$ is not ruled out by observations. This parameter region is relevant because fractional cosmology gives a power-law solution without matter, which is accelerated for $\mu>2$. We present a fractional origin model that leads to an accelerated state without appealing to $\Lambda$ or dark energy.
\end{abstract}


%%%%%%%%%%%%%%%%%%%%%%%%%%%%%%%%%%%%%%%%%%
\date{\today}
\maketitle
%%%%%%%%%%%%%%%%%%%%%%%%%%%%%%%%%%%%%%%%%%
\section{Introduction} 
In fractional calculus, the~classical derivatives and integrals of integer order are generalized to derivatives and integrals of arbitrary (real or complex) order~\cite{monje2010fractional, Tarasov2013,bandyopadhyay2014stabilization,padula2014advances,herrmann2014fractional,tarasov2019applications,klafter2012fractional,malinowska2015advanced,lorenzo2016fractional}. Fractional derivatives have attracted increasing attention because they universally appear as empirical descriptions of complex social and physical phenomena. Fractional calculus applications have grown enormously in recent years because these operators have memory and are more flexible in describing the dynamic behavior of phenomena and systems using fractional differential equations, while the description with integer order differential equations uses local operators and~they are limited in the order of differentiation to a constant. Consequently, the~resulting models must be  sufficiently precise in~many cases~\cite{west2021fractional}. Research into fractional differentiation is inherently multi-disciplinary and has applications across various disciplines, for~example, fractional quantum mechanics and gravity for fractional spacetime~\cite{Calcagni:2010bj, Calcagni:2009kc} and fractional quantum field theory~\cite{Lim:2006hp, LimEab+2019+237+256, VargasMoniz:2020hve, Moniz:2020emn, Rasouli:2021lgy, Jalalzadeh:2021gtq}. Such frameworks have been essential in understanding complex systems in classical and quantum regimes~\cite{rami2009fractional,el2009fractional, El-Nabulsi:2010wwu,el2011fractional,rami2012glaeske,adjm/1351864732, El-Nabulsi:2013hsa,el2015fractional,El-Nabulsi:2015szp,el2018finite,el2020generalized}.

Regarding the classical regime, fractional derivative cosmology has been established by two methods: (i) The last-step modification method is the simplest one, in~which the corresponding fractional field equations replace the given cosmological field equations for a specific model. (ii) The first-step modification method can be considered a more fundamental methodology. In~this method, one starts by establishing a fractional derivative geometry. More concretely, the~variational principle for fractional action is applied to establish a modified cosmological model.

Fractional calculus has recently been explored to address problems related to cosmology in~\cite{Roberts:2009ix, Vacaru:2010fb, Vacaru:2010wj, Vacaru:2010wn, Shchigolev:2010vh, Jamil:2011uj, Shchigolev:2012rp, Debnath2012, El-Nabulsi:2012wpc, El-Nabulsi:2013mwa, El-Nabulsi:2013mma, Debnath2013, Shchigolev:2013jq, Calcagni:2013yqa, Shchigolev:2015rei, Rami:2015kha, El-Nabulsi:2015szp, El-Nabulsi:2016dsj, Calcagni:2016ofu, Calcagni:2016azd, El-Nabulsi:2017vmp, El-Nabulsi:2017jss, Calcagni:2017via, Calcagni:2019ngc,  Calcagni:2020tvw, Calcagni:2020ads, Calcagni:2021ipd, Calcagni:2021aap, Shchigolev:2021lbm, Jalalzadeh:2022uhl, Landim:2021ial, Landim:2021www, Garcia-Aspeitia:2022uxz, Micolta-Riascos:2023mqo}. For~example, in~\citep{Roberts:2009ix,Shchigolev:2010vh}, the~Riemann curvature and the Einstein tensor are defined as usual, but~now with dependence on the $\mu$ fractional parameter. Then, it is possible to write  a fractional analogous to the Einstein field equation through the expression $G_{\alpha \beta}(\mu)=8 \pi G T_{\alpha \beta}(\mu)$, where $G_{\alpha \beta}(\mu)$ is the Einstein tensor in fractional calculus and~$G$ is the Newton gravitational constant. These studies correspond to the last-step modification method, as~mentioned above. Modifications to several astrophysical and cosmological events can be studied based on the last equation. For~example, a~fractional theory of gravitation for fractional spacetime has been developed in \citep{Vacaru:2010fb, Vacaru:2010wj}. Non-holonomic deformations to cosmology lead to new classes of cosmological models, which have been studied in \citep{Vacaru:2010wn, Shchigolev:2021lbm}.

In reference~\cite{Garcia-Aspeitia:2022uxz}, a~joint analysis using data from cosmic chronometers and type Ia supernovae was performed. This comparison with observational tests was used to find best-fit values for the fractional order of the derivative. These methods are a robust scheme for investigating the physical behavior of cosmological models~\cite{Hernandez-Almada:2020uyr, Leon:2021wyx, Hernandez-Almada:2021rjs, Hernandez-Almada:2021aiw, Garcia-Aspeitia:2022uxz} and can be used in new contexts, such as in~\cite{Micolta-Riascos:2023mqo}, where dynamical systems and  phase spaces were used to analyze fractional cosmology for different matter contents, obtaining a late-time accelerating~cosmology.

Undeniably, the~late-time acceleration in the Universe expansion is one of the most challenging topics in modern~cosmology. 

Since 1998, when the independent projects High-z Supernova Search Team~\cite{SupernovaSearchTeam:1998fmf} and Supernova Cosmology Project~\cite{SupernovaCosmologyProject:1998vns} obtained results that suggested this behavior in the Universe, the~type Ia supernovae (SNe Ia) data have become  a definitive proof to study this era of the Universe and the transition between the decelerated expansion phase and the accelerated one. However, the latter  is usually model dependent~\cite{Moresco:2016mzx}. In~this sense, the~observational Hubble parameter data (OHD), also known as cosmic chronometers, have become a fundamental data test, complementary to the SNe Ia data, to~study the Universe's expansion rate in a model-independent way. Finally, using these cosmological data, we estimate the free parameters $(\alpha_0, \mu)$. 

This research's main objective is to investigate open problems in gravity and cosmology. Therefore, we estimate the free parameters using cosmological data. The~analysis of the type Ia supernovae (SNe Ia) data and the observational Hubble parameter data (OHD), also known as cosmic chronometers, and~the joint analysis of   SNe Ia data + OHD lead to best-fit values for the free parameters calculated at the $1\sigma$, $2\sigma$ and $3\sigma$ confidence levels (CLs). On~the other hand, these best-fit values are used to calculate the age of the Universe, the~current deceleration parameter (both at the $3\sigma$ CL) and~the current matter density parameter at $1\sigma$ CL. Finding that the Universe is roughly twice as old as the one of $\Lambda$CDM is a distinction of fractional cosmology; apart from that, it leads to an accelerated state without appealing to $\Lambda$ or dark energy. We confirm that this result, which is in disagreement with the value obtained with globular clusters with~a value of $t_0=13.5^{+0.16}_{-0.14}\pm 0.23$ \citep{Valcin:2021}, is a distinction of  fractional cosmology. This result also agrees with the analysis performed in~(\cite{Garcia-Aspeitia:2022uxz}, {\mbox{Section~5, page~4817}}). 
Despite the discrepancy between the age of the Universe and that determined bY globular clusters, it is essential to highlight that fractional cosmology  contributes to the solutions to other problems associated with the $\Lambda$CDM model, for~example,~late-time acceleration without dark energy, as~we explained before. In this sense, the~non-inclusion of a cosmological constant (CC)  or some DE can alleviate some other problems related to these components. One of these problems is the so-called CC problem, in~which the observational value of the CC differs between $60$ and $120$ orders of magnitude compared with the value anticipated by particle physics~\mbox{\cite{Weinberg:1988cp, Carroll:1991mt, Sahni:1999gb, Peebles:2002gy, Padmanabhan:2002ji}}. Another problem related to the DE is the coincidence problem, which stipulates that, currently, we are living in an extraordinary epoch in the cosmic evolution, in which DM and DE densities are of the same order of magnitude, with~a fine-tuning problem associated with the context of the $\Lambda$CDM model~\cite{Velten:2014nra, Sadjadi:2006qp, Zlatev:1998tr}.

Another issue that fractional cosmology can possibly alleviate is the Hubble tension. Measurements of the Hubble parameter at the current time, $H_{0}$, exhibit a discrepancy of $5\sigma$ between the observational value obtained from the Hubble Space Telescope (HST) \cite{Riess:2021jrx} and~the one inferred from Planck CMB~\cite{Planck:2018vyg}. The~first corresponds to model-independent measurements, while the second depends on the $\Lambda$CDM model. According to~\mbox{\citet{Riess:2019cxk}}, observational issues such as the $H_{0}$ tension are strong evidence that physics beyond the $\Lambda$CDM model is required. 
Therefore, and~following this line, a possible alternative to solve this tension  considers extensions beyond $\Lambda$CDM (see \citet{DiValentino:2021izs} for a review). In~\cite{Garcia-Aspeitia:2022uxz},   some results were discussed related to the $H_0$ tension in the context of fractional cosmology, and   a trend of $H_{0}$ to the value obtained by SH0ES \citep{Riess:2019cxk} at current times was observed,~in agreement with Planck's value \citep{Planck:2018vyg} for $z\lesssim1.5$. However, the~$H_0$ tension is not fully resolved in the region $1.5<z<2.5$.

Moreover, prospects for the density perturbation growth and cosmological structure formation could be described in this context. For~example, in~\cite{Basilakos:2019dof}, it was shown that in the $\Lambda$CDM cosmology, the~perturbations do not change the stability of the late-time attractor of the background equations and~the system still results in the dark-energy-dominated de Sitter solution, with~a dark matter era  transition  and a growth index of $\gamma\approx 6/11$. Here, $\gamma$ is defined through the relation $d \ln \delta_m/d \ln a \approx \Omega_m^{\gamma}$, where $\delta_m$ is the matter contrast and~$\Omega_m$ is the~fractional energy density of matter. This result for the linear growth rate,  $d \ln \delta_m/d \ln a \approx \Omega_m^{\gamma}$, was corrected to $d \ln \delta_m/d \ln a \approx \Omega_m^{\frac{6}{11}} -\frac{1}{70}(1-\Omega_m)^{\frac{5}{2}}$ in~\cite{Alho:2019jho}. In fractional cosmology, the~dimensionless energy density of dust matter depends on $t$ through $\Omega_{m, \mu}=\Omega_m \times t^{-(\mu-1)}$; thus, the~growth index $\gamma$ in the matter-dominated solution should be different to $6/11$. In~addition, it could be exciting to investigate and try fractional cosmology within a very early universe, fitting  data and determining the impact of the fractional derivative term on primordial nucleosynthesis. Fractional cosmology ingredients may enhance inflation, raising the question whether it is possible that issues such as the cosmic no-hair conjecture, isotropization, etc., could be solved~\cite{wald1983asymptotic, kitada1993cosmic, Barrow:1984zz,
Maeda:1988zx, Cotsakis:1993er, Capozziello:1996vp, Bruni:2001pc}.

According to previous statements, it is essential to see if fractional calculus or fractional cosmology can well-describe the  observational data. Then, we can perform more sophisticated calculations to describe the late-time Universe or the very early Universe. Therefore, one can argue that the Universe can be better described with a fractional derivative, not just to fit the data but also to describe the fundamental dynamics, highlighting the demand for new physics. These approaches can help to understand the Universe's acceleration with the mathematical background of fractional calculus. The~mathematical richness generated by  the corrections due to the fractional index $\mu$ of the fractional derivative can resolve the previous problems in future studies. Indeed, we can examine these topics in a forthcoming series of manuscripts with applications in inflation and dark energy models, investigating the physical implications and producing observational constraints.

In this paper, we investigate exact solutions of cosmological interest in fractional cosmology.
In particular, we study the cosmological applications of power-law solutions of the type $a=(t/t_0)^{\alpha_0}$, where $\alpha_0=t_0 H_0$ is the current age parameter. Additionally, given $\mu$,  the~order of Caputo's fractional derivative, and~$w$, the equation of state (EoS) of matter, one must impose two compatibility conditions which allow particular solutions to  $(\mu, w)$
 Moreover, we are interested in an exact solution that gives the general solution of the system. For~this purpose,  one can solve the Riccati equation independent of the EoS, where the solution for the scale factor is a combination of power-law solutions. Additionally, we investigate if the solutions take account of the current late-time acceleration.

The paper is organized as follows. Section~\ref{Sect:II} discusses the basics of the fractional variational approach to cosmology and presents the cosmological equations for a perfect fluid. In~Section~\ref{Sect:III}, we comment on the crucial difference between fractional and standard cosmology; that is, we obtain late-time acceleration without adding a cosmological constant, quintessence scalar field or~other exotic fluids as compared to standard cosmology. In~Section~\ref{Sect:III.1},  we consider a model with cold dark matter to present a specific realization of these possibilities and~we interpret the fractional modification as dark energy in Section~\ref{Sect:III.2}.   Section~\ref{Sect:IV} is devoted to finding exact solutions for the Hubble factor in this scenario. They correspond mainly to power-law solutions for the scale factor and a combination of power-law functions. In Section~\ref{Sect:V}, we provide a precise scheme to find approximated analytical solutions to aid in the asymptotic analysis. A discussion is presented in Section~\ref{Sect:V.1}. We solve Bernoulli's equation using differential inequalities and asymptotic expansions to estimate $H(z)$ in redshift. A~physical discussion of the results is presented in Section~\ref{Sect:VIII}. In~Section~\ref{Sect:VII}, a~joint analysis using OHD and type Ia supernovae data is performed. This comparison with observational tests was used to find best-fit values for the fractional order of the derivative and the current age parameter $\alpha_0$. Section~\ref{Sect:IX} is the conclusion.

\section{Fractional Action~Integral}
\label{Sect:II}

Recently, a~wide range of definitions of fractional derivatives \citep{Uch:2013}, such as the Riemann--Liouville derivative (RLD) and~the Caputo derivative (CD), among~others, have been used in many~applications.

\subsection{Some Fractional~Derivatives}

The RLD with $\mu \geq 0$ for $f(t)$ is defined by
\begin{eqnarray}
D_{t}^{\mu} f(t) & = \Gamma(n-\mu)^{-1} \frac{d^{n}}{d t^{n}} \int_{c}^{t} \frac{f(\tau)}{(t-\tau)^{\mu-n+1}} d \tau, \label{EQ2}
\end{eqnarray}
where $n=[\mu]+1$ and $\mu \in (n-1,n)$.

Note that the main parameter of fractional calculus is given by $\mu$, recovering standard calculus when $\mu\rightarrow 1$.

The Caputo left derivative is defined as
\begin{align}
 & { }^{C} D_{t}^{\mu} f(t) =    \Gamma(n-\mu)^{-1} \int_{c}^{t} \frac{\frac{d^{n}}{d \tau^{n}} f(\tau)}{(t-\tau)^{\mu-n+1}} d \tau, 
\end{align}
where $n=\left\{\begin{array}{cc}
     [\mu] +1& \mu \notin \mathbb{N}\\
     \mu &  \mu \in \mathbb{N}
\end{array} \right.$.
In fractional calculus, we now have the following relation (see \citep{Uch:2013}) for the case of more than one derivative:
\begin{equation}
D_{t}^{\mu} \left[D_{t}^{\beta} f(t)\right]=D_{t}^{\mu+\beta} f(t)-\sum_{j=1}^{n} D_{t}^{\beta-j} f(c+) \frac{(t-c)^{-\mu-j}}{\Gamma(1-\mu-j)},
\end{equation}
or in other words $D_{t}^{\mu} D_{t}^{\beta} f(t) \neq D_{t}^{\mu+\beta} f(t)$, if~not all derivatives $D_{t}^{\beta-j} f(c+)$  are equal to zero at $c$. 
Additionally, the~fractional derivative of the Leibniz rule \citep{Uch:2013} reads as
\begin{equation}
D_{t}^{\mu}[f(t) g(t)]=\sum_{k=0}^{\infty} \frac{\Gamma(\mu+1)}{k ! \Gamma(\mu-k+1)} D_{t}^{\mu-k} f(t) D_{t}^{k} g(t),
\end{equation}
having the usual when $\mu=n\in \mathbb{N}$.

\subsection{Frational Action-like Variational~Problems}

Within the first-step modification method, one procedure uses the fractional variational approach developed in~\cite{El-Nabulsi:2005oyu, El-Nabulsi:2007lla, El-Nabulsi:2007wgc, El-Nabulsi:2008oqn,Roberts:2009ix, 10.5555/1466940.1466942} with~the following fractional action integral,
\begin{align}
 S(\tau)
   & = \frac{1}{\Gamma(\mu)}\int_0^\tau \mathcal{L}\left(\theta, q_i(\theta), \dot{q}_i(\theta), \ddot{q}_i(\theta)\right)(\tau-\theta)^{\mu-1} d\theta, \label{GenCFL}
\end{align}
\noindent 
where $\Gamma(\mu)$ is the Gamma function, $\mathcal{L}$ is the Lagrangian, $\mu$ is the constant fractional parameter and~$\tau$ and $ \theta$ are the observers and~intrinsic time, respectively, and the action integral depends on second order derivatives of the generalized coordinates $q_i$.

Variation in \eqref{GenCFL} with respect to $q_i$ leads to the Euler--Poisson equations \cite{10.5555/1466940.1466942}:
\begin{align}
&   \frac{\partial \mathcal{L}\left(\theta, q_i(\theta), \dot{q}_i(\theta), \ddot{q}_i(\theta)\right)}{\partial q_i}   -\frac{d}{d \theta} \frac{\partial \mathcal{L}\left(\theta, q_i(\theta), \dot{q}_i(\theta), \ddot{q}_i(\theta)\right) }{\partial \dot{q}_i}  + \frac{d^2}{d \theta^2} \frac{\partial \mathcal{L}\left(\theta, q_i(\theta), \dot{q}_i(\theta), \ddot{q}_i(\theta)\right) }{\partial \ddot{q}_i}  \nonumber \\
& =  \frac{1-\mu}{\tau-\theta} \left(\frac{\partial \mathcal{L}\left(\theta, q_i(\theta), \dot{q}_i(\theta), \ddot{q}_i(\theta)\right)}{\partial \dot{q}_i}   -2 \frac{d}{d \theta} \frac{\partial  \mathcal{L}\left(\theta, q_i(\theta), \dot{q}_i(\theta), \ddot{q}_i(\theta)\right)}{\partial \ddot{q}_i} \right)  \nonumber \\
& - \frac{\left(1-\mu\right)\left(2-\mu\right)}{\left(\tau-\theta\right)^2} \frac{\partial  \mathcal{L}\left(\theta, q_i(\theta), \dot{q}_i(\theta), \ddot{q}_i(\theta)\right)}{\partial \ddot{q}_i}. \label{EP}
 \end{align}

\subsection{Applications to~Cosmology}
\label{Sect:III}

In cosmology, it is assumed that the flat Friedmann provides the geometry of spacetime Lema\^{i}tre--Robertson--Walker (FLRW) metric:
\begin{equation}
  ds^2=-N^2 (t) dt^2+ a^2(t)(dx^2+ dy^2+dz^2),  \label{FLRWm}
\end{equation}
where  $a(t)$ denotes the scale factor and $N(t)$ is the lapse function.
This result is based on Planck's observations~\cite{Planck:2018vyg}.

For the metric \eqref{FLRWm}, the~Ricci' scalar  depends on the second derivatives of $a$ and first derivatives of $N$ and reads
\begin{equation}
R(t)=    6\Bigg(\frac{\ddot{a}(t)}{a(t) N^2(t)} +\frac{{\dot{a}}^2(t)}{a^2(t) N^2(t)}-\frac{\dot{a}(t) \dot{N}(t)}{a(t) N^3(t)}\Bigg). \label{Rscalar}
\end{equation}

Consider the point-like action  integral
\begin{align}
 S (\tau)  = \int_0^\tau \left[\frac{R(\theta)}{2}+ {L}(\theta)\right] a^3(\theta) N(\theta) d\theta, \label{CFL0}
\end{align}
where $R(\theta)$ is the Ricci scalar \eqref{Rscalar}. 
In cosmology, the~Einstein--Hilbert Lagrangian density is related to the Ricci scalar. Generically, one takes integration by parts, such that a total derivative is removed for the action, and~the derivatives $\ddot{a}(t)$ and $\dot{N}(t)$ are eliminated. We will use a fractional version of the Lagrangian \eqref{CFL0}; thus, we do not follow the standard procedure and keep the higher order derivatives and use  formulation \eqref{GenCFL} leading to  the Euler--Poisson Equation~\eqref{EP}.  We use fractional variational calculus with classical and Caputo derivatives.

For simplicity, we consider units in which $ 8\pi G=c=1$ and assume a perfect fluid for the matter content of the Universe,
where ${L}(\theta)=-\rho_{0} a(\theta)^{-3(1+w)}$ for~$w\neq -1$ is the usual matter Lagrangian of a perfect fluid as in
~\cite{Wald, Carroll, carroll2004spacetime} and contains integer-order derivatives in the Lagrangian. We consider the transition to the effective fractional action  used in~\cite{Garcia-Aspeitia:2022uxz}, i.e.,
\begin{align}
 S_{{\text{eff}}} (\tau)
   & = \frac{1}{\Gamma(\mu)}\int_0^\tau \Bigg[\frac{R(\theta)}{2}  +\mathcal{L}(\tau, \theta)\Bigg](\tau-\theta)^{\mu-1} a^3(\theta) N(\theta) d\theta, \label{CFL}
\end{align}
\noindent
where $\Gamma(\mu)$ is the Gamma function, $\mathcal{L}(\tau, \theta)={L}(\theta) (\tau -\theta )^{-(\mu -1) (w+1)}$, which recovers the usual matter Lagrangian of a perfect fluid as $\mu\rightarrow 1$ 
\cite{Wald, Carroll, carroll2004spacetime}, $\mu$ is the constant fractional parameter and~$\tau$ and $ \theta$ are the observers and intrinsic time, respectively \cite{Shchigolev:2010vh}. $w=p/\rho$ is a constant EoS for~matter. 

For a fixed $\tau$, the~expressions
\begin{equation}
    \rho(\theta)=\rho_{0} a(\theta )^{-3(1+w)} (\tau -\theta )^{-(\mu -1) (w+1)}, \label{rho(theta-tau)}
\end{equation}
and
\begin{equation}
p(\theta)=w \rho_{0} a(\theta )^{-3(1+w)} (\tau -\theta )^{-(\mu -1) (w+1)}, \label{p(theta-tau)}
\end{equation} 
define the energy density and the isotropic pressure of the matter fields. Then, \vspace{-9pt}

\begin{align}
  & \dot{\rho}(\theta)=\frac{d}{d\theta}\left[\rho_{0} a(\theta)^{-3(1+w)}(\tau -\theta )^{-(\mu -1) (w+1)}\right]=(w+1) \rho(\theta ) \left(-\frac{3 \dot{a}(\theta )}{a(\theta )}-\frac{\mu -1}{\theta
   -\tau }\right)\nonumber\\
  & = -3 \left(\frac{ \dot{a}(\theta)}{a(\theta )}+\frac{1-\mu}{3(\tau-\theta)}\right) (\rho(\theta)+p(\theta)).\label{cons}
\end{align}

Defining $q_i\in \{N, a~\}$  in \eqref{GenCFL}  for a fixed $\tau$, we have the Lagrangian \vspace{-9pt}

\begin{align}
 & \mathcal{L}\left(\theta, N(\theta), \dot{N}(\theta), a(\theta), \dot{a}(\theta), \ddot{a}(\theta)\right):=  \nonumber \\
 & \frac{3 a(\theta ) \left(N(\theta ) \left(a(\theta ) \ddot{a}(\theta )+\dot{a}^2(\theta
   )\right)-a(\theta ) \dot{a}(\theta ) \dot{N}(\theta )\right)}{ N^2(\theta )}- \rho_{0} N(\theta ) a(\theta )^{-3 w} (\tau -\theta )^{-(\mu -1) (w+1)}.
\end{align}

The Euler--Poisson equations \eqref{EP} obtained after varying the action \eqref{CFL} for $q_i\in \{N, a~\}$ lead to the field equations  \vspace{-9pt}

\begin{align}
&\left(\frac{\dot{a}(\theta )}{a(\theta )}\right)^2 + \frac{(1-\mu)}{(\tau-\theta )} \frac{\dot{a}(\theta )}{a(\theta )}= \frac{1}{3}\rho_{0} a(\theta )^{-3(1+w)} (\tau -\theta )^{-(\mu -1) (w+1)}, \label{FEQ1}
\\
& \frac{\ddot{a}(\theta )}{a(\theta )}+\frac{1}{2}\left(\frac{\dot{a}(\theta )}{a(\theta )}\right)^2+ \frac{(1-\mu)}{(t-\theta )} \frac{\dot{a}(\theta )}{a(\theta )}+\frac{(\mu -2) (\mu -1)}{2 (\tau-\theta )^2}=-\frac{1}{2} \rho_{0} w a(\theta )^{-3(1+w)}
   (\tau -\theta )^{-(\mu -1) (w+1)}. \label{REQ1}
\end{align}
\noindent

Here, we have substituted the lapse function $N=1$ after the~variation.

To designate the temporary independent variables, the~rule $(\tau, \theta) \mapsto (2t,t)$ is applied, where new cosmological time $t$ \cite{Shchigolev:2010vh} is used, where the dots denote these derivatives. Furthermore, the~Hubble parameter is $H\equiv\dot{a}/a$. Hence, Equations \eqref{FEQ1} and \eqref{REQ1} and the conservation Equation \eqref{cons} can be written as
\begin{align}
& \dot{H}(t)+\frac{(1-\mu ) H(t)}{t}+\frac{3 H(t)^2}{2}+\frac{(\mu -2)
   (\mu -1)}{2 t^2}=-\frac{1}{2}  p(t) \label{Raych1},\\
& H(t)^2+\frac{(1-\mu ) H(t)}{t}= \frac{1}{3} \rho (t), \label{Fried1}\\
& \dot{\rho}(t)= -3 \left(H(t) +\frac{(1-\mu)}{3 t}\right)(p(t)+ \rho(t)) \label{Cons} 
\end{align}
where expressions \eqref{rho(theta-tau)} and 
\eqref{p(theta-tau)} are transformed to
\begin{equation}
    \rho(t)=\rho_{0} \left[a(t )^{-3} t^{-(\mu -1)}\right]^{1+w}, \;\text{and}\;
p(t)=w \rho_{0}  \left[a(t )^{-3} t^{-(\mu -1)}\right]^{1+w},
\end{equation} 
which defines the energy density and the isotropic pressure of the matter fields in cosmological~time. 

\subsection{Some Cosmological~Solutions}

From  \eqref{Fried1} and assuming $p=0 \; \left( w =0\right) $, the~Hubble parameter is
\begin{equation}
H\left( t\right) =\frac{\mu -1}{2t}\left[ 1+\sqrt{1+\frac{4}{3}\rho \left(
t\right) \left( \frac{t}{\mu -1}\right) ^{2}}\right].  \label{IVA}
\end{equation}%

We have considered the positive root because we are interested in expanding~universes.

To understand the self-accelerating behavior of $H$, let us assume that there is no matter, i.e., $\rho=0$ and~$\mu>1$. Then, from~\eqref{Fried1},
\begin{equation}
H\left( t\right) =\frac{\mu -1}{t}\implies a\left( t\right) =\left( 
\frac{t}{t_{0}}\right) ^{\mu -1}.\label{3}
\end{equation}%

Henceforth,
\begin{equation}
H\left( t\rightarrow \infty \right) \rightarrow 0, \quad %
a\left( t\rightarrow \infty \right) \rightarrow \infty,  \quad  \text{if}\; \mu >1 \label{4}
\end{equation}%
and the deceleration parameter can be expressed as $1+q=-\dot{H}/H^{2}$. Therefore,
\begin{align}
q & = -1+\frac{1}{\mu -1},  \label{5} \\
\; \text{and}\;\mu &> 1\Longrightarrow q>-1,  \label{6}
\end{align}%
where the usual case $q=-1$, corresponding to a cosmological constant $\Lambda$, is~excluded.

The case $\rho=0$ can also be interpreted as a fluid whose energy density
quickly vanishes with evolution. The~asymptotic solution \eqref{3} was 
examined in detail using a dynamical systems analysis in reference~\cite{Garcia-Aspeitia:2022uxz}, and~its properties are summarized in Table~\ref{Tab2}.
The table summarizes the asymptotic behavior for $\mu>1$ when the energy density of matter tends to zero. Hence, even in the absence of matter, fractional cosmology gives a power-law solution $a(t)= \left(t/t_0\right)^{\mu-1}$, which is accelerated for $\mu>2$. This is a crucial difference to standard cosmology, where we must add a cosmological constant, quintessence scalar field, or~other exotic fluids to accelerate the~expansion. 

\begin{table}[H]

\caption{\label{Tab2} Asymptotic solution $P_4$,  examined  in reference~\cite{Garcia-Aspeitia:2022uxz}. 
}
    \setlength{\tabcolsep}{2.5mm}
    \resizebox{1\textwidth}{!}{%
    \begin{tabular}{ccccccc}
    \toprule 
\textbf{Label} & $\mathbf{\Omega_{\text{m}}}$  & \textbf{H} & \textbf{q} & \textbf{Acceleration?}  & \textbf{Stability}& \textbf{Scale Factor} \\\midrule
$P_4$ &  $0$ & $\frac{\mu -1}{t}$ & $-\frac{\mu -2}{\mu -1}$ & Accelerated  ($\mu>2$) & Sink ($\mu >2$) & Power law\\
&&&& Decelerated ($1<\mu <2$) &  Source ($\mu <{7}/{4}$) & $a(t)= \left(t/t_0\right)^{\mu-1}$\\ 
&&&& & Saddle (${7}/{4}<\mu <2$) & \\\toprule 
\end{tabular}}
\end{table}

Solving algebraically Equations \eqref{Raych1}--\eqref{Cons} for $\dot{H}$, $\dot{\rho}$ and $\rho$, we obtain
\begin{align}
   & \dot{H} = -\frac{1}{2} p+\frac{(\mu -1) H}{t}-\frac{3}{2} H^2-\frac{(\mu
   -2) (\mu -1)}{2 t^2}, \label{NewRaych}\\
   & \dot{\rho}= -\frac{3 (\mu -1)^2 H }{ t^2}+\frac{12
   (\mu -1) H ^2}{t}- 9 H ^3 +p \left(\frac{\mu -1}{t}-3 H\right), \label{NewCons}\\
   & \rho= 3 H^2-\frac{3 (\mu -1) H}{ t}. \label{NewFried}
\end{align}

In~General Relativity (GR), we have the flat Friedmann--Lemaître--Robertson--Walker metric; the main equations are the 
Friedmann constraint and conservation equation,
\begin{align}
3H^{2} & = \rho, \label{a} \\
\dot{\rho}+3H\left( \rho +p\right)& = 0.  \label{b}
\end{align}%

Using  \eqref{a} and \eqref{b},
we obtain
\begin{align}
2\dot{H}=-\left( \rho +p\right).%
  \label{d}
\end{align}%

Now, using \eqref{a} and \eqref{d} 
we re-obtain \eqref{b}. That is, we have three equations, two of them independent. 
However, as~we discussed before, \cite{Micolta-Riascos:2023mqo} studied Equations \eqref{Raych1}--\eqref{Cons}, and~using a similar procedure as in GR, we obtain a new equation (see Equation~\eqref{NewP}) instead of showing that two out of three equations are~independent.

 By demanding that \eqref{Fried1} is conserved in time, i.e.,
\begin{equation}
    \frac{d}{d t} \left[H(t)^2+\frac{(1-\mu ) H(t)}{t}-  \frac{1}{3}\rho (t)\right]=0,
\end{equation}
we calculate the corresponding derivatives and substitute them into \eqref{NewRaych}--\eqref{NewFried} to obtain
\begin{equation}
    \frac{(\mu -1) (t ( t p-3 H (t H+2 \mu -6))+3 (\mu -2) (\mu -1))}{6t^3}=0.
\end{equation}

This equation is an identity for $\mu=1$ as expected in standard cosmology. However, for~$\mu\neq 1$, we acquire the new relation for the pressure of the fluid:
\begin{equation}
   p(t) = \frac{6 (\mu -3) H(t)}{t}+3 H^2(t)-\frac{3 (\mu -2)(\mu -1)}{ t^2}. \label{NewP}
\end{equation}

Using a similar procedure as in GR, we obtain a new Equation~\eqref{NewP} instead of showing that two out of three equations are independent. This characteristic of fractional cosmology leads to some restrictions of the matter fields in the Universe that were explored in~\cite{Micolta-Riascos:2023mqo} for different matter~fields.

Replacing the expression of $p$ defined by \eqref{NewP} into \eqref{NewRaych} and \eqref{NewCons}, we obtain
\begin{align}
    & \dot{H}= -\frac{2 (\mu -4) H}{t}-3 H^2+\frac{(\mu -2) (\mu -1)}{t^2}, \label{Riccati}\\
    & \dot{\rho}= \frac{3 (\mu -1) (4 \mu -11) H}{ t^2}-\frac{3 (\mu -13) H^2}{t}-18 H^3 -\frac{3 (\mu -2) (\mu -1)^2}{ t^3}.
\end{align}

The previous results are valid for any ideal gas~source. 

Moreover, following  references~\cite{Garcia-Aspeitia:2022uxz, Micolta-Riascos:2023mqo}, the~system can be extended by including several matter sources in Equations \eqref{FEQ1} and \eqref{REQ1}. After~performing algebra, and~using  $8\pi G=1$, the~following Raychaudhuri equation (with $N(t)=1$) is obtained:
\begin{equation}
   \dot{H}+\frac{(\mu -1) H}{2 t}+\frac{(\mu -2) (\mu -1)}{2 t^2}=- \frac{1}{2}\sum_i (p_i+\rho_i), \label{IntegrableH}
\end{equation}
along with the Friedmann equation
\begin{equation}
    H^2+\frac{(1-\mu)}{t}H=\frac{1}{3}\sum_i\rho_i. \label{FriedmannFrac}
\end{equation}
\indent Furthermore, the continuity equation leads to
\begin{equation}
    \sum_i\left[\dot{\rho}_i+3\left(H+\frac{1-\mu}{3t}\right)(\rho_i+p_i)\right]=0,\label{CE}
\end{equation}
where $\rho_i$ and $p_i$ are the density and pressure of the $i$th matter component, respectively, and the~sum is over all species, e.g.,~matter, radiation, etc. Note that when $\mu=1$ in Formula \eqref{FriedmannFrac} and Formula \eqref{CE}, the~standard cosmology without $\Lambda$ is recovered, which by itself does not produce an accelerated expanding universe.

Using the equation of state $p_i=w_i\rho_i$, where $w_i \neq -1$ and are constants, we have
\begin{align}
       & \sum_i (1+w_i)\rho_i \left[\frac{\dot{\rho}_i}{ (1+w_i) \rho_i} + 3\frac{\dot{a}}{a}+ \frac {1-\mu}{t} \right] = \sum_i (1+w_i)\rho_i \frac{\mathrm{d}}{\mathrm{d}t}\left[ \ln \left( {\rho_i}^{ 1/(1+w_i)} a^3 t^{1-\mu}\right)\right].
\end{align}

Assuming separate conservation equations for each species and integrating for each ${\rho_i}$, we have the following solution:
\begin{align} 
\rho_{i}(t)= \rho_{0i} a(t)^{-3 (1+w_i)} \left(t/t_0\right)^{(\mu-1)(1+w_i) }, \label{EQ16}
\end{align}
where $a(t_0)=1$,  $t_0$ is the age of the universe and $\rho_{0i}$ is the current value of the energy density of the $i$th species. 
Therefore, by~substituting \eqref{EQ16} into \eqref{FriedmannFrac}, we have
\begin{align}
   & H^2+\frac{(1-\mu)}{t}H =\frac{8\pi G}{3}\sum_i \rho_{0i} a^{-3 (1+w_i)}\left(t/t_0\right)^{(\mu-1)(1+w_i)}. \label{EH}
\end{align}

Note that for $\mu\neq 1$, the~modified continuity Equation \eqref{CE} provides the condition
\begin{align}
    \frac{8 \pi G}{3} \sum_i p_i=\frac{2 (\mu -3) H}{t}+H^2-\frac{(\mu -2) (\mu -1)} {t^2}. \label{Newpressure}
\end{align}

{Combining these results with \eqref{IntegrableH}
and \eqref{FriedmannFrac}, we have the Riccati equation~\eqref{Riccati}. This equation generically appears in fractional cosmology, independent of the matter content. Therefore, in~the following, we consider only one matter source.}

Comparing with other fractional formulations, according to~\cite{Shchigolev:2013jq, Shchigolev:2015rei}, and~assuming $\Lambda=0$ and using  $8\pi G \Gamma(\mu) = 1$ for simplicity, we obtain the field equations
\begin{align}
& 2\dot{H}+3H^{2}+2\frac{\left( 1-\mu \right) }{t}H+\frac{\left( 1-\mu \right)
\left( 2-\mu \right) }{t^{2}}  = -t^{1-\mu } P,   \label{f2}\\ 
& 3H^{2}+3\left( 1-\mu \right) \frac{H}{t} = t^{1-\mu }\varrho,  \label{f1} \\
& \dot{\varrho}+3H\left(\varrho + P\right)= 0, \label{f3}
\end{align}%
where $\varrho$ and $P$ are the bare dark matter energy density and pressure, respectively.
Equations \eqref{f2}--\eqref{f3} are equivalent to \eqref{Raych1}--\eqref{Cons} under~the scaling
\begin{equation}
(\rho,  p)= t^{1-\mu } (\varrho, P).
\end{equation}

Now, we consider a constant EoS,
\begin{equation}
 P=w \varrho. 
 \end{equation}
 
Then, one obtains,
\begin{equation}
w =-1+\frac{2}{3}\left[ 1+q(t) -\frac{\left(
\mu -1\right) }{2tH(t)}-\frac{\left( \mu -1\right) \left( \mu -2\right) }{%
2\left( tH(t)\right) ^{2}}\right] \left( 1-\frac{\mu -1}{tH(t)}\right) ^{-1}. 
\label{f4}
\end{equation}%

Then, for~$\mu =1$ we have
\begin{align}
3H^{2}& = \rho, \label{f5} \\
2\dot{H} +3H^{2}& =  -p \implies 2\dot{H}=-\left( \rho +p\right),  \label{f6} 
\end{align}
and
\begin{align}
q=\frac{1}{2}\left[ 1+3\left( \frac{p}{\rho }\right) \right] =\frac{1}{2}\left( 1+3w \right) \implies  w =\frac{1}{3}\left( 2q-1\right). 
\label{f7}
\end{align}%

Furthermore, we recover~GR.

From Equations \eqref{f1} and \eqref{f2} and~$\mu\neq 1$, it follows that
\begin{equation}
2\dot{H}-\frac{\left( 1-\mu \right) }{t}H+\frac{\left( 1-\mu \right) \left(
2-\mu \right) }{t^{2}}=-t^{1-\mu }\left( 1+w \right) \varrho.  \label{f8}
\end{equation}%

Replacing \eqref{f1}, we recover \eqref{f4}.

Setting  $%
w =0$, from~\eqref{f4} we obtain
\begin{align}
q\left( t\right)  & = \frac{1}{2}+\left( \frac{\mu -1}{tH(t)}\right) \left( 
\frac{\left( \mu /2-1\right) }{tH(t)}-1\right).   \label{f9}
\end{align}%

That is, a~``correction''  to the usual {\text{CDM}} ($\mu=1$, $w =0$, $q=1/2$).

Using \eqref{f9} and~the relation $q=-1-\dot{H}/H^2$, we obtain the equation for $H$ as
\begin{equation}
\dot{H}=\frac{(\mu -1) H}{t}-\frac{3}{2} H^2-\frac{(\mu -2) (\mu -1)}{2 t^2}.  \label{f11}
\end{equation}

For $\mu=1$, we recover the usual {\text{CDM}} ($w =0$, $q=1/2$),
\begin{equation}
    H(t)= \frac{2 H_0}{3 H_0 (t-t_0) +2}, \; q= \frac{1}{2}.
\end{equation}

For $\mu=2$, the~solution of \eqref{f11} is
\begin{equation}
 H(t)=   \frac{4 H_0 t}{3 H_0 (t-t_0) (t+t_0)+4 t_0}, \; q(t)=\frac{1}{2}- \frac{1}{t H}= -\frac{1}{4} -\frac{t_0}{H_0 t^2}+\frac{3 t_0^2}{4 t^2}.
\end{equation}

We will search for solutions considering relativistic matter/radiation.
Setting  $w =1/3$, from~\eqref{f4}, we obtain
\begin{align}
q\left( t\right)  & = 1+ \frac{(\mu -2) (\mu -1)}{2 t^2
   H(t)^2}-\frac{3 (\mu -1)}{2 t H(t)}.   \label{Radf9}
\end{align}

Using \eqref{Radf9} and~the relation $q=-1-\dot{H}/H^2$, we obtain the equation for $H$ as
\begin{equation}
\dot{H}=  \frac{3 (\mu -1) H}{2 t}-2 H^2-\frac{(\mu -2) (\mu -1)}{2 t^2}.  \label{Radf11}
\end{equation}

For $\mu=1$, we recover the usual radiation ($w =1/3$, $q=1$),
\begin{equation}
    H(t)= \frac{H_0}{2H_0 (t-t_0) +1}, \; q=1.
\end{equation}

For $\mu=2$, the~solution of \eqref{Radf11} is
\begin{equation}
 H(t)=\frac{5 H_0 t^{3/2}}{4 H_0 \left(t^{5/2}-t_0^{5/2}\right)
   +5 t_0^{3/2}},
\end{equation} 
and
\begin{align}
 q(t) & = \frac{(\mu -2) (\mu -1) t_0^3
   (5-4 H_0 t_0)^2}{50H_0^2 t^5}-\frac{(\mu -1) (8
   \mu -31) t_0^{3/2} (4 H_0 t_0-5)}{50 H_0t^{5/2}} \nonumber \\
   & +\frac{1}{25} \left(8 \mu ^2-54 \mu +71\right).
\end{align}

Another case of interest is when $w =-1$ (quasi-vacuum matter, according to~\cite{Shchigolev:2013jq, Shchigolev:2015rei}). 

The equation for $H$ becomes
\begin{equation}
   \dot{H}=-\frac{(\mu -1) H}{2 t}-\frac{(\mu -2) (\mu -1)}{2 t^2}.
\end{equation}

The solution to $H\left( t\right) $ turns out to be
\begin{equation}
H\left( t\right) =H_0 \left(\frac{t}{t_0}\right)^{-\frac{\mu-1}{2}}+ \frac{(\mu -2) (\mu -1)}{(3-\mu) t}\left[1-  \left(\frac{t}{t_0}\right)^{\frac{3-\mu }{2}}\right].  \label{f16}
\end{equation}%

According to \eqref{f16},
\begin{align}
q\left( t\right) =-1+\Bigg[\frac{H_0 (\mu -1) \left(\frac{t}{t_0}\right)^{\frac{1}{2}-\frac{\mu }{2}}}{2 t} & - \frac{(\mu -2)
   (\mu -1) \left(1-\left(\frac{t}{t_0}\right)^{\frac{3}{2}-\frac{\mu }{2}}\right)}{(\mu -3) t^2} \nonumber \\
   & +\frac{(\mu
   -2) (\mu -1) \left(\frac{t}{t_0}\right)^{-\frac{\mu }{2}-\frac{1}{2}}}{2 t_0^2}\Bigg] H(t)^{-2}.  \label{f17}
\end{align}

For $0<\mu<3$, we have
\begin{equation}
q\left( t\rightarrow \infty \right) \rightarrow -1.  \label{f18}
\end{equation} 

Hence, we obtain a late-time de Sitter solution without including a cosmological constant. 

In summary, fractional cosmology allows for an accelerated expansion without adding exotic fluids to the model. Therefore, we now consider a model with cold dark matter to present a specific realization of these~possibilities.  

\subsection{Model with Cold Dark~Matter}
\label{Sect:III.1}
Assume
\begin{equation}
    H\left( t\right) =\frac{\alpha}{t},\label{12}
\end{equation}
where $\alpha$ is a constant and $\mu>1$. If $\alpha=\mu-1$, we recover Equation \eqref{3}. 

The conservation equation for  matter \eqref{Cons}, for~cold dark matter ($p_{\text{CDM}}=w_{\text{CDM}}\rho _{\text{CDM}}$ and $w_{\text{DM}}=0$), takes the form
\begin{equation}
\dot{\rho}_{\text{CDM}}+3H\left( 1+\frac{1-\mu }{3Ht}\right) \rho_{\text{CDM}}=0,\label{8}
\end{equation}%
when then reduces to
\begin{equation}
\dot{\rho}_{\text{CDM}} +  \frac{(3 \alpha -\mu +1) {\rho}_{\text{CDM}}}{t}=0.
\end{equation}

Hence, we have for the matter energy density,
\begin{align}
\rho _{\text{CDM}} & = \rho_{CDM\left( 0\right) }(t/t_0)^{^{-3 \alpha +\mu -1}}.
\end{align}% 

Choosing
\begin{equation}
 \alpha = \frac{\mu +1}{3}, \label{zeta}
\end{equation}
we obtain
\begin{equation}
\rho _{\text{CDM}}  = \rho_{CDM\left( 0\right) }(t/t_0)^{-2}.
\end{equation}

Then, from~Equation \eqref{NewFried}, we obtain
\begin{align}
\rho_{CDM\left( 0\right) }= \frac{2 (2-\mu) (\mu +1)}{3 t_0^2}. \label{13}
\end{align}% 

Using the  redshift parameter  $1+z=1/a$, we obtain
\begin{align}
H\left( t\right) & = \frac{\alpha}{t}\implies  a\left(
t\right) =\left( \frac{t}{t_{0}}\right) ^{\alpha},  \label{17} \\
\; \text{and}\;1+z & = \frac{1}{a}\Longrightarrow \left( \frac{t}{t_{0}}%
\right)=\left( 1+z\right) ^{-\frac{1}{\alpha}},  \label{18}
\end{align}%
then,
\begin{equation}
\rho _{\text{CDM}}\left( z\right) =\rho _{\text{CDM}}\left( 0\right) \left(
1+z\right) ^{\frac{2}{\alpha}}\; \text{and}\; H\left( z\right)
=H_0 \left( 1+z\right) ^{\frac{1}{\alpha}}, 
\label{19}
\end{equation}%
where $\alpha$ is defined by \eqref{zeta}. 
Comparing with GR, where the EoS $w_{\text{eff}}$ is defined through
\begin{equation}
\rho _{\text{CDM}}\left( z\right) =\rho _{\text{CDM}}\left( 0\right)  (1+z)^{3(1+w_{\text{eff}})},\label{20}
\end{equation}%
we have
\begin{equation}
w_{\text{eff}} =  -1 + \frac{2}{3 \alpha}= -1+\frac{2}{\mu +1} \; \text{and}\; 
q  = -1+\frac{1}{\alpha}= -1+\frac{3}{\mu +1}. \label{18c}
\end{equation}%

Similarly to GR, we have the usual relation   $q= \frac{1}{2} \left(1 + 3 w_{\text{eff}}\right)$. 
Therefore, in~fractional cosmology, we  have an acceleration ($\ddot{a}>0, q<0$) as is present in GR when the effective fluid has $w_{\text{eff}}<-1/3$. 
Hence,
\begin{equation}
\ddot{a}\left( t\right) <0, q>0, w_{\text{eff}}>-1/3 \Longleftrightarrow 1< \mu<2,  \label{19b}
\end{equation}
and
\begin{equation}
\ddot{a}\left( t\right) > 0, q<0, w_{\text{eff}}<-1/3 \Longleftrightarrow \mu >2. \label{19c}
\end{equation}

Finally, we have an accelerated expansion if $\mu >2$, caused by the fractional derivative correction and not by the matter content. That is the  powerful advantage of fractional cosmology over GR. This is consistent as $\rho _{\text{CDM}}\rightarrow 0$ with the asymptotic solution \mbox{$H(t)=\frac{\mu -1}{t}$}, where~ $q=-\frac{\mu -2}{\mu -1}$, which is a power-law solution  $a(t)= \left(t/t_0\right)^{\mu-1}$. It is accelerated if $\mu>2$ and decelerated if $1<\mu <2$, as proven in~\cite{Garcia-Aspeitia:2022uxz}.

\subsection{Interpretation of the Fractional Term as a Dark Energy~Source}
\label{Sect:III.2}
We write  \eqref{Fried1} as
\begin{align}
& 3H^{2} = \rho _{\text{CDM}}+\rho _{\text{frac}},  \label{23} \\
& \; \text{where}\; \rho _{\text{CDM}}\left( z\right) =\rho _{\text{CDM}}\left( 0\right) \left(
1+z\right) ^{\frac{2}{\alpha}}= \frac{2 (2-\mu) (\mu +1)}{3 t_0^2}  \left(
1+z\right) ^{\frac{6}{1+\mu}}, \\
& \; \text{where}\;\rho _{\text{frac}}\left( t\right) = \frac{3\left( \mu -1\right) }{t%
}H \underbrace{\implies}_{\text{using}\; \eqref{12}\; \text{and} \; \eqref{18}} \rho _{\text{frac}}\left( z\right) =\rho _{\text{frac}}\left( 0\right)
\left( 1+z\right) ^{\frac{2}{\alpha }},  \label{24} \\
& \; \text{and}\;\rho _{\text{frac}}\left( 0\right) = \frac{3\left( \mu -1\right) }{%
t_{0}}H_{0}=\frac{\left( \mu -1\right) }{t_{0}H_{0}}3H_{0}^{2}.   \label{25}
\end{align}

Using  $H=d\ln a/dt$ and $1+z=1/a$, we obtain
\begin{equation}
t_{0}H_{0}=\int_{0}^{\infty }\frac{dz}{\left( 1+z\right) E\left( z\right) }.
\label{41}
\end{equation}

Substituting (see Equation \eqref{12})
\begin{equation}
E\left( z\right) =\left( 1+z\right) ^{\frac{1}{\alpha }}= \left( 1+z\right) ^{\frac{3}{1+\mu}}, 
\label{42}
\end{equation}
with $\alpha$ defined by \eqref{zeta}, 
we obtain
\begin{equation}
H_0t_0=(1+\mu)/3. \label{H0t0mu}
\end{equation}

On the other hand, using \eqref{23}, \eqref{24} and \eqref{25}, we obtain
\begin{align}
& E^{2}\left( z\right)  = \frac{3H^{2}\left( z\right) }{3H_{0}^{2}}=\frac{\rho
_{\text{CDM}}\left( z\right) }{3H_{0}^{2}}+\frac{\rho _{\text{frac}}\left( z\right) }{%
3H_{0}^{2}}\Longleftrightarrow E^{2}\left( z\right) =\Omega _{\text{DM}}\left(
z\right) +\Omega _{\text{frac}}\left( z\right),   \label{26} \\
& \implies \Omega _{\text{frac}}\left( 0\right) =1-\Omega _{\text{DM}}\left( 0\right) =%
\frac{\mu -1}{t_{0}H_{0}}= \frac{3(\mu -1)}{\mu+1}\sim 0.744\%, \;  \mu \sim 1.65957,  \label{28}
\end{align}%
where $\Omega_{\text{DM}}\left( 0\right) \sim 0.256\%$. Compare the value $\mu\sim 1.66$, with~the observational tests performed in~\cite{Garcia-Aspeitia:2022uxz} for a flat prior $1< \mu<3$, where the best-fit value for $\mu$ is $\mu^*= 1.71$.

We inspect more the nature of $\rho _{\text{frac}}$ as an effective fluid in GR, i.e.,
\begin{align}
\rho _{\text{frac}} & = \frac{3\left( \mu -1\right) }{t}H,\; q= -1-\frac{\dot{H}}{H^2}
\implies  \dot{\rho}_{\text{frac}} =-H\left( 1+q+\frac{1}{Ht}\right) \rho _{\text{frac}},  \label{32} \\
&\implies \dot{\rho}_{\text{frac}}+3H\left( 1+w _{\text{frac}}\right) \rho
_{\text{frac}}=0,  \label{33}
\end{align}%
where $w_{\text{frac}} = \frac{1}{3}\left(q-2+\frac{1}{\alpha}\right)$. According to \eqref{18c}, we deduce again \eqref{20},
\begin{align}
w_{\text{frac}} & = -1+\frac{2}{3\alpha },  \label{36}
\end{align}%
corresponding to quintessence ($-1<w_{\text{frac}}<-1/3$) if $\mu >2$.

\section{Exact~Solutions}
\label{Sect:IV}

\subsection{First Exact~Solution}

From Equations \eqref{NewFried} and \eqref{NewP}
and defining the effective equation of state $ w:= {p}/{\rho}$, we have
\begin{equation}
    w= \frac{\mu -2}{t H}+\frac{2 \mu -5}{t H-\mu +1}. \label{First_Eq_H}
\end{equation}

Assuming $w\neq -1$ and is a constant, and~by solving \eqref{First_Eq_H} algebraically for $H$, we obtain
\begin{align}
& H_{1,2}(t) = \frac{\alpha_\pm}{t}, \label{solH-1}\\
& \alpha_\pm ={-\frac{6-2 \mu +(1-\mu)w \pm  \sqrt{\mu ^2 \left(w^2+8\right)+w^2-2 \mu  (w (w+2)+18)+4 w+44}}{2 (w-1)}}. \label{apm-1} 
\end{align}

Hence, in~the intervals $-1<w<1, 1<\mu <2$, both $H_{1,2}(t)$ are non-negative. For~ $-1<w<1, \mu>2$, $H_1(t)$ is negative and $H_2(t)$ is positive. For~$\mu\in\{1,2\}$, $H_1(t)$ is~zero. 

The deceleration parameter for each algebraic solution is a constant
\begin{align}
& q_{1,2}(t):= -1-{\dot{H_{1,2}}(t)}/{H_{1,2}^2(t)}=-1+ {1}/{\alpha_\pm},
\end{align}
such that the solutions for the scale factor are~power laws. 

Therefore, upon~physical consideration, we select the one that gives an accelerated Universe. The~deceleration parameter $q_1(t)$ is negative in the parameter region $-1<w<1, 1<\mu <2$, and~we have an accelerated expansion when $H(t)=H_1(t)$ and $(\mu,w) \in [-1,1] \times[1, 2]$. If~we choose the range $-1<w<1, 2<\mu<3$, the~solution $H_1(t)$ becomes nonphysical, and~the one that gives accelerated expansion is $H_2(t)$ because $q_2(t)$ is negative in the parameter region $-1<w<1, 2<\mu <3$.

Substituting \eqref{solH-1}, \eqref{NewFried} and \eqref{NewP} into \eqref{NewRaych} and \eqref{NewCons}, we obtain the compatibility conditions
\begin{align}
   & -2 \alpha_\pm  \mu -3 (\alpha_\pm -3) \alpha_\pm +\mu ^2-3 \mu +2=0, \label{compatibility_1}\\
   &(2 \alpha_\pm -\mu +1) \left(2
   \alpha_\pm  \mu +3 (\alpha_\pm -3) \alpha_\pm -\mu ^2+3 \mu -2\right)=0, \label{compatibility_2}
\end{align}
where $\alpha_\pm$ is defined by \eqref{apm-1} for~the existence of an exact solution. 
We define the current value of $H(t_0)=H_0$ through $H_0 t_0 =\alpha_\pm$, and~$\alpha_\pm$  is interpreted as the age parameter $\alpha= t H$ evaluated at $t_0$ (the current time). 

Therefore, to~obtain solutions, we solve \eqref{compatibility_1}
and \eqref{compatibility_2} simultaneously for $\mu$ and $w$. Upon~physical consideration, we remove the cases with $w=-1$ and $\mu \in\{1,2\}$, and~we assume $-1<w<1$. 

From \eqref{solH-1}, the~definition of $q$ and Equations \eqref{NewFried}, \eqref{NewP} and~\eqref{First_Eq_H}, we have  
\begin{align*}
& a(t)= \left(\frac{t}{t_0}\right)^{\alpha_\pm}, \quad H(t)= \frac{\alpha_{\pm}}{t}, \quad  q(t)=-1+ \frac{1}{\alpha_{\pm}},\\
& p(t)= \frac{3 \left(\alpha_\pm^2+2 \alpha_\pm (\mu -3)-(\mu -2) (\mu -1)\right)}{ t^2}, \quad  \rho(t)= \frac{3 \alpha_\pm (1-\mu+\alpha_\pm )}{ t^2}, 
\end{align*} and
\begin{align}
& w:= {p(t)}/{\rho(t)}=\frac{\mu }{\alpha_{\pm}}
+\frac{2 \alpha_{\pm}-3}{1-\mu+\alpha_{\pm}}-\frac{\alpha_{\pm}+2}{\alpha_{\pm}}, \label{alpha_w}
\end{align}
where we set the conditions  $a'(t_0)=H_0,$ and fix the current value of the scale factor to $a(t_0)=1$. 

For simplicity, let us assume that the source is dust, with~$p=0$ ($w=0$). 
Then, we have $\alpha_\pm= 3-\mu +\epsilon\sqrt{\mu  (2 \mu -9)+11}$, where $\epsilon=\pm 1$, which makes \eqref{alpha_w} an~identity. 

Hence, we have
\begin{align}
&  H_{1,2} (t)= \frac{3-\mu +\epsilon\sqrt{\mu  (2 \mu -9)+11} }{t},  \\
& \rho(t)=\frac{3 \left(\mu - \epsilon\sqrt{\mu  (2 \mu -9)+11} -3\right) \left(2 \mu
   -\epsilon\sqrt{\mu  (2 \mu -9)+11}  -4\right)}{ t^2}.
\end{align}

Substituting in \eqref{NewRaych} and \eqref{NewCons},
we obtain the compatibility conditions (replace $w=0$ in~\eqref{compatibility_1}
and \eqref{compatibility_2}),
\begin{align}
  & -3 (\mu  (2 \mu -9)+11)+\sqrt{\mu  (2 \mu -9)+11} (4 \mu -9) \epsilon +2=0,\\
  & -(17 \mu -39)
   (\mu  (2 \mu -9)+11)+6 \mu +6 (\mu  (2 \mu -9)+11)^{3/2} \epsilon \nonumber \\
   & +(\mu  (12 \mu -55)+59)
   \sqrt{\mu  (2 \mu -9)+11} \epsilon -14=0.
\end{align}

For  $\epsilon=-1$, the~two conditions are simultaneously satisfied only for $\mu\in\{1,2\}$. 

 For $\epsilon=1$, the~two conditions are simultaneously satisfied only for $\mu\in\{5/2, 7/2\}$. Since we are interested in the case where $\mu\notin\{1,2\}$ and an expanding Universe (\mbox{$H>0$}), the~parameter that gives the physical solution $H(t)=H_2(t)$ is $\mu=5/2$. Substituting these values for $w$ and $\mu$, we obtain
\begin{equation}
\left( p(t),  \rho(t), H(t)\right)  \rightarrow \left(0,0,\frac{3}{2 t}\right) \implies a(t)= a_0 t^{3/2}.
\end{equation}

That means if we fix the equation of state $w$, there are specific values that $\mu$ has to satisfy to obtain an exact~solution.

\subsection{Second Exact~Solution}

One can also solve  a Riccati equation for $\mu\neq 1$:
\begin{align}
    \dot{H}(t)=\frac{(\mu -1) (3 w+2) H(t)}{2 t}-\frac{3}{2} (w+1) H(t)^2-\frac{(\mu -2) (\mu -1)}{2
   t^2} \label{Rccati2}
\end{align}
which follows from substituting $p=w \rho$ into \eqref{NewRaych} and removing $\rho$ using \eqref{NewFried}. 
We obtain the exact solution
\begin{align}
& H(t)= \frac{1}{3 t (w+1)}    \Bigg[\mu+\beta+\frac{3}{2} (\mu -1) w -\frac{2 b_1  t_0^{\beta } \beta}{t^\beta+b_1 t_0^{\beta } }\Bigg], \label{exactRicatti2}
    \end{align}
where
\begin{equation}
    b_1=  \frac{2 (\beta +\mu )-6 H_0 t_0 (w+1)+3 (\mu -1) w}{2 \beta +6 H_0 t_0 (w+1)-2 \mu -3 \mu  w+3 w}
\end{equation} is an integration constant,  $H_0$ is the current value of $H$ at $t=t_0$ and~\begin{equation}
\beta=\frac{1}{2}\sqrt{-4 (\mu  (2 \mu -9)+6)+9 (\mu -1)^2 w^2+24 (\mu -1) w}.
\end{equation}

Substituting in \eqref{NewRaych} and \eqref{NewCons},
we obtain the compatibility conditions\vspace{-9pt}

\begin{align}
& -4 \beta  (\beta -4 \mu +9)+\frac{8 \beta ^2 (w-1) t^{2 \beta
   }}{\left(t^{\beta }+b_1 t_0^{\beta }\right){}^2}-3 (\mu -1) (3 \mu -13) w^2  +w \left(4\beta  (5 \mu -12)-8 \mu ^2+42 \mu -6\right)  +24  \nonumber \\
   & -\frac{8 \beta  t^{\beta } (-\beta +4 \mu +w (\beta +5 \mu-12)-9)}{t^{\beta }+b_1 t_0^{\beta}}=0,
\\
   &t^{2 \beta } \left(4 \beta  (\beta
   +4 \mu -9)+3 (\mu -1) (3 \mu -13) w^2+w \left(4 \beta  (5 \mu -12)+8 \mu ^2-42 \mu
   +6\right)-24\right) \nonumber \\
   & +2 b_1 t^{\beta } t_0^{\beta } \left(w \left(4 \beta ^2+\mu ^2 (9
   w+8)-6 \mu  (8 w+7)+39 w+6\right)-24\right) \nonumber \\
   & +b_1{}^2 t_0^{2 \beta } \left(4 \beta 
   (\beta -4 \mu +9)+3 (\mu -1) (3 \mu -13) w^2+w \left(\beta  (48-20 \mu )+8 \mu ^2-42 \mu
   +6\right)-24\right)=0.
   \end{align}

These compatibility conditions have to be satisfied for all $t$, such that $b_1=0$. Then, \eqref{exactRicatti2} becomes \begin{align}
& H(t)= \frac{1}{3 t (w+1)}    \Bigg[\mu+\beta+\frac{3}{2} (\mu -1) w\Bigg], \label{exactRicatti22}
    \end{align}
and the compatibility conditions are  \vspace{-9pt}

\begin{align}
   & -4 \beta  (\beta +4 \mu -9)-3 (\mu -1) (3 \mu -13) w^2+w \left(\beta  (48-20 \mu )-8
   \mu ^2+42 \mu -6\right)+24=0, \label{compatibility_21}\\
   &4 \beta  (\beta +4 \mu -9)+3 (\mu -1)
   (3 \mu -13) w^2+w \left(4 \beta  (5 \mu -12)+8 \mu ^2-42 \mu +6\right)-24=0. \label{compatibility_22}
\end{align}

Therefore, to~obtain solutions, we solve \eqref{compatibility_21}
and \eqref{compatibility_22} simultaneously for $\mu$ and $w$.
Upon physical consideration, we remove the cases with $w=-1$ and $\mu \in\{1,2\}$, and~we assume $-1<w<1$. 

As before, we choose dust matter ($p=0, w=0$). Then,
\begin{align}
    & H(t)= \frac{\mu + \sqrt{(9-2 \mu ) \mu -6} \left(1-\frac{2 b_1}{b_1+t^{\sqrt{(9-2 \mu ) \mu
   -6}}}\right)}{3 t}.
\end{align}

As before, two compatibility conditions are satisfied only for $b_1=0$ and $\mu\in\{5/2, 7/2\}$.

For $b_1=0$ and $\mu=5/2$, we have the physical solution
\begin{equation}
\left( p(t),  \rho(t), H(t)\right)  \rightarrow \left(0,0,\frac{3}{2 t}\right) \implies a(t)= a_0 t^{3/2}.
\end{equation}

For $b_1=0$ and $\mu=7/2$, we have the nonphysical solution
\begin{equation}
 \left( p(t),  \rho(t), H(t)\right)  \rightarrow    \left(0,-\frac{9}{2 t^2},\frac{3}{2 t}\right).
\end{equation}
  
As in the previous section, if~we impose the equation of the state of the fluid as dust, this fixes the values of $\mu$ to $5/2$. 

\subsection{General~Solution} \label{Gen_Sol_H}  
In this section, we are interested in an exact solution that gives the general solution of the system. For~this purpose,  one can solve the Riccati equation
\eqref{Riccati} independent of the EoS. That has the $H(t)$ solution defined by
\begin{align}
    H(t)=\frac{1}{3 t }\left[\frac{9-2 \mu +r}{2}-\frac{c r \alpha_0^r}{c  \alpha_0^r+ (H_0  t)^r}\right],\label{V}
\end{align}  
where
\begin{align}
c & =  \frac{-2 \mu +r-6 \alpha_0+9}{2 \mu +r+6 \alpha_0-9}, 
\; \text{and}\;r  = \sqrt{8 \mu  (2 \mu -9)+105}, 
\end{align}%
and for the current time $\alpha_0=H_{0}t_0$, where $t_0$ is the value of $t$ today. $H_0$ is the current value of the Hubble factor, $\alpha_0$, for which we obtain the best-fit~values. 

That is the exact solution for $H(t)$ studied in~\cite{Micolta-Riascos:2023mqo} (see an analogous case in~\cite{Shchigolev:2012rp}, Equation~(36), and~in~\cite{Garcia-Aspeitia:2022uxz}, Equation~(24)). 
In this case, expressions \eqref{NewFried} and \eqref{NewP} are used to calculate $\rho(t)$ and $p(t)$. Substituting all the expressions in the system \eqref{NewRaych}--\eqref{NewFried} leads to identities.
There is an arbitrary constant of integration and~the equations are identically satisfied (no compatibility equations are required); thus, this is the general solution of the system. This result is generic since it does not require specifying the EoS. Hence, Equation \eqref{V} gives a one-parameter family of solutions that gives a complete solution and is independent of the matter~content. 

Defining the dimensionless time variable $\tau=H_{0}t$ such that $\alpha_0=H_{0}t_0$ and~$\xi=\tau/\alpha_0=t/t_0$,  by~definition, the~current value of $\xi$ is $\xi_0=1$ and the~expressions  become \vspace{-9pt}
\begin{small}
\begin{align}
 & a(\xi)=\xi^{\frac{1}{6} (-2 \mu -r+9)} \sqrt[3]{\frac{{c+\xi^r}}{{(c+1)}}}, \label{(65)}\\
 & z(\xi)= -1+\xi^{\frac{1}{6} (2 \mu +r-9)} \sqrt[3]{\frac{{(c+1)}}{{c+\xi^r}}}, \label{ztau} \\
 & E(\xi)= \frac{H(\xi)}{H_0} =\frac{1}{3 \alpha_0 \xi } \left[\frac{9-2 \mu +r}{2}-\frac{c r}{c +\xi^r}\right], \label{(66)}
\\
 & p(\xi) = \frac{H_0^2 \left(2 (4 \mu -9) r \left(\xi^{2 r}-c^2\right)+r^2 \left(\xi^r-c \right)^2-7 (4 \mu (2 \mu -9)+45) \left(c +\xi^r\right)^2\right)}{12 \alpha_0^2 \xi^2 \left(c +\xi^r\right)^2}, \label{(68)}\\
 & \rho(\xi) = \frac{H_0^2 \left(-2 (5 \mu -12) r \left(\xi^{2 r}-c^2 \right)+r^2 \left(\xi^r-c \right)^2+(2 \mu -9) (8 \mu -15)
 \left(c  +\xi^r\right)^2\right)}{12 \alpha_0^2 \xi^2 \left(c +\xi^r\right)^2}, \label{(69)}
\\
 & q(\xi)=-\frac{c^2 (2 \mu +r-9) (2 \mu +r-3) +2 c \left(4 \mu ^2-24 \mu +5 r^2+27\right) \xi^r+(-2 \mu +r+3) (-2 \mu +r+9) \xi^{2 r}}{\left((-2 \mu +r+9) \xi^r-c (2 \mu +r-9)\right)^2}, \label{(67)}\\
 & w_{\text{eff}}(\xi)=\frac{2 (4 \mu -9) r \left(\xi^{2 r}-c^2\right)+r^2 \left(\xi^r-c \right)^2-7 (4 \mu  (2 \mu -9)+45)
   \left(c +\xi^r\right)^2}{\left((-2 \mu +r+9) \xi^r-c (2 \mu +r-9)\right) \left((-8 \mu +r+15) \xi^r-c (8 \mu +r-15)\right)}, \label{(70)}\\
 & \Omega_{\text{m}}(\xi)= \frac{(-8 \mu +r+15) \xi^r-c (8 \mu +r-15)}{(-2 \mu +r+9) \xi^r-c (2 \mu +r-9)} \label{(71)}.
\end{align}

 Taking the limit $\xi \rightarrow \infty$, we have 
  \begin{align*}
& \lim_{\xi \rightarrow \infty} z(\xi)=-1, \quad 
\lim_{\xi \rightarrow \infty} a(\xi)=\infty, \quad 
\lim_{\xi \rightarrow \infty} E (\xi)=0, \quad 
\lim_{\xi \rightarrow \infty} p(\xi)=0, \quad 
\lim_{\xi \rightarrow \infty} \rho(\xi)=0, 
\\
& \lim_{\xi \rightarrow \infty} q(\xi)
=\frac{-13-2 (\mu -4) \mu +\sqrt{8 \mu  (2 \mu -9)+105}}{2 (\mu -2) (\mu -1)}, 
\\
& \lim_{\xi \rightarrow \infty}  w_{\text{eff}} (\xi)= \frac{-7+\sqrt{8 \mu  (2 \mu -9)+105}}{4 (\mu -1)}, \quad \lim_{\xi \rightarrow \infty}   \Omega_{\text{m}}  (\xi) =\frac{5-\sqrt{8 \mu  (2 \mu -9)+105}}{2 (\mu -2)}, 
\end{align*}
\end{small}
and \begin{equation}
\lim_{\xi \rightarrow \infty} \alpha (\xi) =  \frac{1}{6} \left(9-2 \mu +\sqrt{8 \mu  (2 \mu -9)+105}\right)\geq 0,
\end{equation}
where $\alpha(t)=t H$ is the age~parameter.
 
The main difficulty of this approach is the need to invert \eqref{ztau} to obtain $\xi$ as a function of $z$ because the data are in terms of redshift, which is impossible using analytical tools. After~all, the~equation is a rational one. However, the~variable $\xi$ can be used as a parameter instead of $z$ in the parametric~representation.

\subsection{Asymptotic~Analysis}
\label{Sect:V}
The following is a precise scheme which does not require inverting \eqref{ztau}.

By introducing the logarithmic independent variable $s= -\ln(1+z)$, with~$s\rightarrow -\infty $ as $z\rightarrow \infty$, $s\rightarrow 0$ as $z\rightarrow 0$ and~ $s\rightarrow \infty $ as $z\rightarrow -1$, and~defining the age parameter as  $\alpha = t H$, we obtain the initial value problem
\begin{align}
   & \alpha '(s)= 9-2 \mu -3 \alpha (s)+\frac{(\mu -2) (\mu -1)}{\alpha (s)}, \label{eq(134)}\\
   &t'(s)={t(s)}/{\alpha(s)}, \label{eq(135)}\\
   &  \alpha(0)=t_0 H_0, t(0)=t_0. \label{eq(136)}
\end{align}

Equation \eqref{eq(134)} gives a one-dimensional dynamical system. The~equilibrium points are 
$T_1: \alpha = \frac{1}{6} \left(9-2 \mu -\sqrt{8 \mu  (2 \mu -9)+105}\right)$, that satisfies $\alpha>0$ for $1<\mu <2$, and~$T_2: \alpha= \frac{1}{6} \left(9-2 \mu +\sqrt{8 \mu  (2 \mu -9)+105}\right)$,  that satisfies $\alpha>0$ for $\mu\in\mathbb{R}$. 

The eigenvalue of $T_1$ is $-3- 36 (\mu -2) (\mu -1)\Big/\left(2 \mu +\sqrt{8 \mu  (2 \mu -9)+105}-9\right)^2>0$ for  $1<\mu <2$. Hence,  $T_1$ is a source whenever it~exists.  

The eigenvalue of $T_2$ is  $-3- 36 (\mu -2) (\mu -1)\Big/\left(-2 \mu +\sqrt{8 \mu  (2 \mu -9)+105}+9\right)^2<0$  for $\mu\in\mathbb{R}$. Hence,  $T_2$ is always a sink. It has asymptotic behavior for large $s$, which is consistent with~\cite{Micolta-Riascos:2023mqo}, in~which the attractor solution has an asymptotic age parameter
\begin{align}
    \lim_{t\rightarrow \infty} t H= \frac{1}{6} (9-2 \mu +r). \label{eq135}
\end{align}

We introduce the parameter $\epsilon_0$ such that
\begin{equation}
 \epsilon_0= \frac{1}{2}\lim_{t\rightarrow \infty} \left(\frac{t_0 H_0 - t H}{t H}\right), \quad \alpha_0=  \frac{1}{6} \left(9 -2 \mu +\sqrt{8 \mu  (2 \mu -9)+105}\right)(1+ 2 \epsilon_0), \label{alpha_0}
\end{equation} 
where $\epsilon_0$ is a measure of the limiting value of the relative error in the age parameter $t H$ when it is approximated by $t_0 H_0$. 
When $\epsilon_0=0$,  $\alpha_0=\frac{1}{6} (-2 \mu +r+9)$, which implies $c=0$ in~\eqref{ztau}
and \eqref{(66)}. Thus, we obtain the leading term
\begin{align}
    & E(z)= (1+z)^{\frac{6}{ (9-2 \mu +r)}}. \label{(142)}
\end{align}

We obtain the exact value of $E(z)$ by integrating the initial value problem numerically
\begin{align}
& E'(z)=\frac{E(z) \tau (z) (3 E(z) \tau (z)+2 \mu -8)-(\mu -2) (\mu -1)}{(z+1)E(z) \tau (z)^2}, \; E(0)=1, \label{syst_1}\\
& \tau '(z)=-\frac{1}{(1+z)E(z)}, \; \tau (0)=\frac{1}{6} (2\epsilon_0 +1) (9-2 \mu +r). \label{syst_2}
\end{align}

\subsection{Approximated Analytical~Solution}
\label{Sect:V.1}

Substituting $\alpha=e^{-3 s} y ^{\frac{1}{2}}$
into \eqref{eq(134)}, it is transformed into the
following equation:
\begin{align}
& \frac{d y}{d s}= 2 (\mu -2) (\mu -1) e^{6 s}+2 (9-2 \mu) e^{3 s} y^{\frac{1}{2}}. \label{eq(136)b} 
\end{align}

Let $m$ be the solution of the following Bernoulli equation
\begin{equation}
 \frac{dm}{ds}=2 (9-2 \mu) e^{3 s} m^{\frac{1}{2}}.   \label{eqm}
\end{equation}

The solutions of Bernoulli's Equation~\eqref{eqm} are three:
\begin{equation}
\label{(Eq:139)}
m(s)=\left\{\begin{array}{cc}
0 & \text{such that}\;   m(0)=0\\
 \frac{1}{9} \left[(9-2 \mu)(1-  e^{3 s}) -3 m_0^{\frac{1}{2}}\right]^2 & \text{such that}\; m(0)=m_0 \\
\frac{1}{9}\left[(9-2 \mu)(1-  e^{3 s}) +3 m_0^{\frac{1}{2}}\right]^2  &  \text{such that}\; m(0)=m_0
\end{array}
\right..
\end{equation}

Taking the difference term by term of \eqref{eq(136)b} and
\eqref{eqm}, we obtain
\begin{align}
& \frac{d y}{d s}- \frac{dm}{ds}= 2 (\mu -2) (\mu -1) e^{6 s}+2 (9-2 \mu) e^{3 s} \left[ y^{\frac{1}{2}}-  m^{\frac{1}{2}}\right].\label{(EQ129)}
\end{align}

Assuming $y=m+n$, where $n\geq 0$ is the remainder in the approximation of $y$ by $m$, and~considering the following inequality
\begin{equation}\left(m+n\right) ^{\frac{1}{2}}\leq m^{\frac{1}{2}}+n^{\frac{1}{2}}\label{ineq}
\end{equation}
for $m\geq 0$ and $n\geq 0$, from~\eqref{(EQ129)}, we then obtain the differential inequality
\begin{equation}
\frac{d n}{d s}\leq 2 (\mu -2) (\mu -1) e^{6 s}+2 (9-2 \mu) e^{3 s} n^{\frac{1}{2}}. \label{eqN}
\end{equation}

Suppose that
$n = A^2 e^{6 s}$, where~$A$ is to be determined.
Then, Equation \eqref{eqN} leads to
\begin{equation}
 3 A^2 \leq  (\mu -2) (\mu -1) + (9-2 \mu)  A. \label{inequality}
\end{equation}

The equality occurs
at  $A$ values of
\begin{align}
  &  A_-= \frac{1}{6} \left(9-2 \mu -\sqrt{8 \mu  (2 \mu -9)+105}\right),   A_+=\frac{1}{6}\left(9-2 \mu +\sqrt{8 \mu  (2 \mu -9)+105}\right),
\end{align}
which are the $\alpha$ values of the equilibrium points $T_1$ and $T_2$ of the one-dimensional dynamical system \eqref{eq(134)}.

We choose the third solution of Bernoulli's equation in \eqref{(Eq:139)}. Hence,  we obtain an approximation of $y$, given by
\begin{align}
    y_{\text{approx}}=  \frac{1}{9} \left[(9-2 \mu)(1-  e^{3 s}) + 3 m_0^{\frac{1}{2}}\right]^2  + A^2 e^{6 s}.
\end{align}

Then, we obtain an approximation of $\alpha$, given by
\begin{align}
    \alpha_{\text{approx}}(s)= \left[ \frac{1}{9} \left((9-2 \mu)(1-  e^{3 s}) + 3 m_0^{\frac{1}{2}}\right)^2 e^{-6 s} + A^2 \right]^{\frac{1}{2}}. 
\end{align}

Calculating the limit
\begin{align}
    \lim_{s \rightarrow +\infty}   \alpha_{\text{approx}}(s)=  \frac{1}{3} \sqrt{9 A^2+(9-2 \mu )^2},
\end{align}
and imposing the equality with \eqref{eq135}, we have
\begin{align}
A^2 & = \frac{1}{9} \mu  \left(\mu -\sqrt{8 \mu  (2 \mu -9)+105}+9\right)+\frac{1}{2} \sqrt{8 \mu  (2 \mu -9)+105}-\frac{23}{6}, \label{(EQ140)}
\end{align}
which is non-negative and satisfies  inequality \eqref{inequality} for  $2\leq \mu \leq \frac{1}{10} \left(63+\sqrt{849}\right)\lesssim 9.21376$.  Moreover, $A \in [0, A_+]$ for  $2\leq\mu \leq3.25162$ or~$7.59791\leq\mu \leq9.21376$. 

To calculate $m_0$, we use the condition $\alpha_{\text{approx}}(0)=\alpha_0:= t_0 H_0$. 
Hence, $m_0= \alpha_0^2-A^2$.

In terms of redshift, we have
\begin{align}
    \alpha_{\text{approx}}(z)= \left[ \frac{1}{9} \left((9-2 \mu)\left(1-  (1+z)^{-3}\right) + 3 \left(\alpha_0^2-A^2\right)^{\frac{1}{2}}\right)^2 (1+z)^6 + A^2 \right]^{\frac{1}{2}}, 
\end{align}
where $A$ is defined by \eqref{(EQ140)}.

For Equation \eqref{eq(135)}, we obtain
\begin{equation}
  t^{-1} \frac{dt}{ds}=\left[ \frac{1}{9} \left((9-2 \mu)(1-  e^{3 s}) + 3 m_0^{\frac{1}{2}}\right)^2 e^{-6 s} + A^2 \right]^{-\frac{1}{2}},
\end{equation}
with the solution given by
\begin{align}
t(s)& =t_{0}\exp \left( \bigints_{0}^{s} \left[ \frac{1}{9} \left((9-2 \mu)(1-  e^{3 \zeta}) + 3 m_0^{\frac{1}{2}}\right)^2 e^{-6 \zeta} + A^2 \right]^{-\frac{1}{2}} d\zeta\right).
\end{align}

Now, considering  an asymptotic expansion of the integral for large values of$s$, 
where $m_0= \alpha_0^2-A^2$ and~$A$ is defined by \eqref{(EQ140)}, we obtain
\begin{align}
t(s)&  \simeq t_{0}e^{\bigints_{0}^{s} \frac{3}{\sqrt{9 A^2+(9-2 \mu )^2}}  d\zeta} =t_{0} e^{\frac{3 s}{\sqrt{9 A^2+(9-2 \mu )^2}}}. \label{eqAsymptotict}
\end{align}

Consequently,
\begin{align}
&H(s)\simeq \frac{H_0}{\alpha_0}\left[ \frac{1}{9} \left((9-2 \mu)(1-  e^{3 s}) + 3 m_0^{\frac{1}{2}}\right)^2 e^{-6 s} + A^2 \right]^{\frac{1}{2}} e^{-\frac{3 s}{\sqrt{9 A^2+(9-2 \mu )^2}}}.
\end{align}

We have substituted an asymptotic expansion of the integral for large $s$ given by \eqref{eqAsymptotict}.

Finally,
\begin{align}
& E(z) = \frac{H(z)}{H_0} \simeq E_{\text{approx}}(z)\nonumber \\
& = \frac{1}{\alpha_0} \left[ \frac{1}{9} \left((9-2 \mu)\left(1-  (1+z)^{-3}\right) + 3 m_0^{\frac{1}{2}}\right)^2 (1+z)^{6} + A^2 \right]^{\frac{1}{2}} \left(1+z\right)^\frac{3}{\sqrt{9 A^2+(9-2 \mu )^2}} \label{Ez}
\end{align}
and
\begin{equation}
t(z) \simeq t_{\text{approx}}(z)=t_{0} (1+z)^{-\frac{3}{\sqrt{9 A^2+(9-2 \mu )^2}}}, \label{tz}
\end{equation}
as $z\rightarrow -1$, 
where $\alpha_0$ is defined by \eqref{alpha_0}, 
$A$ is defined by \eqref{(EQ140)} and $m_0= \alpha_0^2-A^2$.

\begin{figure}[H]
  %  \centering
    \includegraphics[scale=0.6]{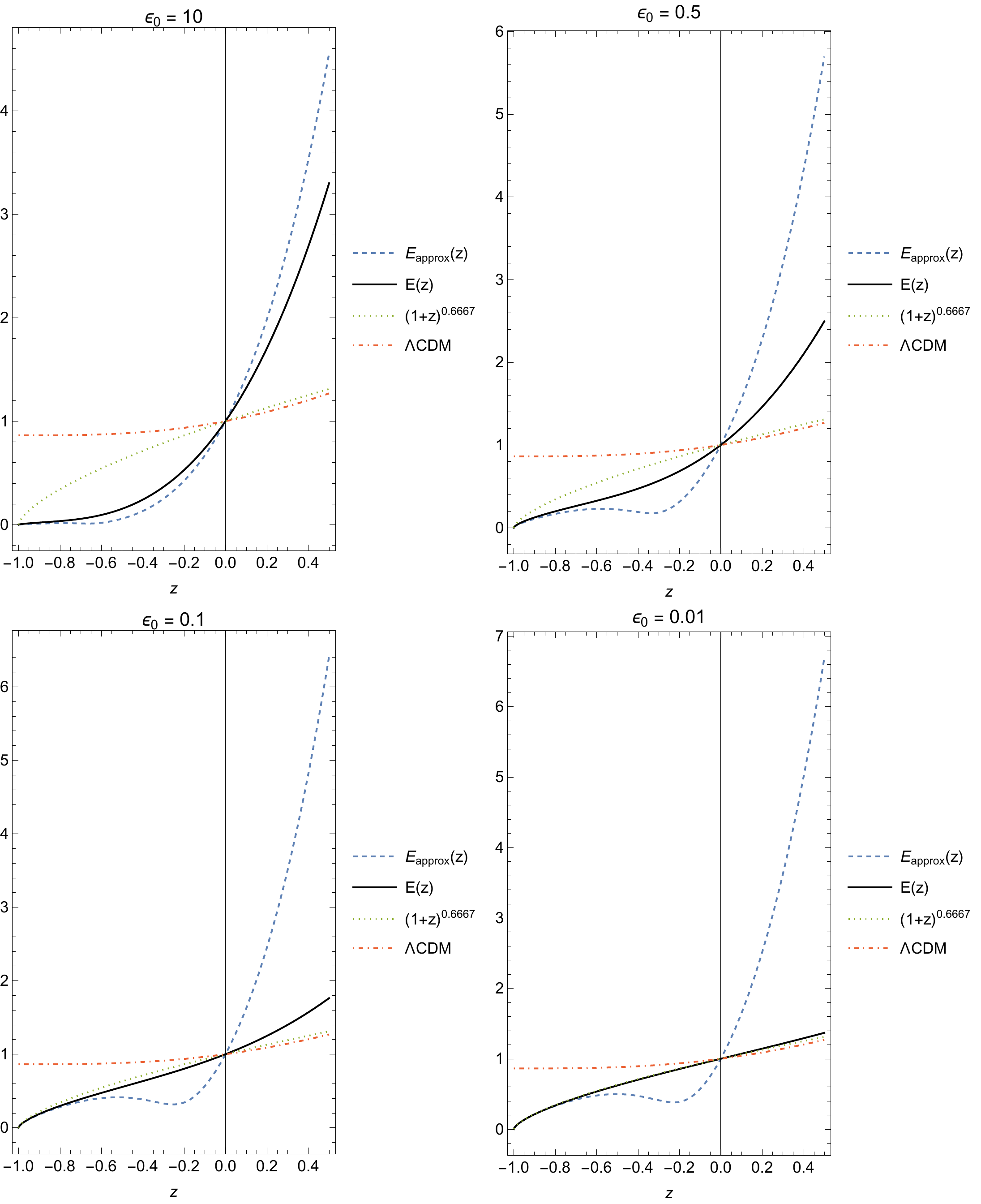}
    \caption{Expression \eqref{Ez} compared the numerical solution of \eqref{syst_1} and \eqref{syst_2}, the~leading term  $ (1+z)^{\frac{6}{ (9-2 \mu +r)}}$,  $E(z)$ ($\Lambda$CDM) for $\mu=2.5$ and different values of $\epsilon_0$. }
    \label{fig:Ez}
\end{figure}
%%%author's request: please keep the figure here. The figure it should appears before it is referenced. Otherwise we have a huge blank space. 
Figure~\ref{fig:Ez} presents Expression \eqref{Ez} compared to the numerical solutions of \eqref{syst_1} and \eqref{syst_2},~leading term  $ (1+z)^{\frac{6}{ (9-2 \mu +r)}}$ and~$E(z)$ of $\Lambda$CDM for $\mu=2.5$ and different values of $\epsilon_0$. The figure  shows that generically, for~ $2\leq \mu \lesssim 9.21376$, there is a good accuracy of the approximation of the exact value $E(z)$ by the asymptotic approximation \eqref{Ez} as $z\rightarrow -1$. 
\subsection{Discussion}
\label{Sect:VIII}
Our analysis shows differences between standard $\Lambda$CDM cosmology-based GR and the fractional version. Given the energy density expression $\rho $ in GR, one can calculate the Hubble parameter through the Friedmann equation. Therefore, we deduce $H$ and~we investigate the cosmological history. Finally, we  consider the existence of dark matter ($w =0$), as $w =-1$ for $\Lambda$ and $-1<w<-1/3$ for quintessence,  as~suggested from the~observations.

If we proceed as before, to~give $w$,~we use  $w=p/\rho $ and then use the equations for $\rho$ and $p$, i.e., Equations \eqref{NewFried} and \eqref{NewP}. With~these expressions, one can calculate $H\left(t\right) $ and $q\left( t\right)$. In~fractional cosmology, the~asymptotic behavior $H\left( t\right) \sim 1/t$ is a characteristic due to the fractional parameter $\mu>1$. 

We  related  the possible matter scenarios, and~according to the relevant discussion on  the EoS $w$, we have two regimes of~interest. 

First, consider $\rho$-like dark matter with the behavior of dark energy, $-1<w <-1/3$, i.e., quintessence. This case was investigated and~the relevant results are summarized in the~following.

In the first approach, we solve  \eqref{First_Eq_H} for $H$, obtaining two solutions $H_{1,2}(t)= \frac{\alpha_\pm}{t}$, where the $\alpha$ parameter is defined by \eqref{apm-1} and takes the values $\alpha_\pm$. 

Therefore, to~obtain solutions, we solved \eqref{compatibility_1}
and \eqref{compatibility_2} simultaneously for $\mu$ and $w$. 
For $\epsilon=-1$, we have the following solutions for Equations \eqref{compatibility_1}
and \eqref{compatibility_2}:
\begin{itemize}
    \item $w= \frac{7+\sqrt{8 \mu  (2 \mu -9)+105}}{4(1- \mu)}\; \text{ if }\; \mu <1 \; \text{or}\; 1<\mu
   <2 \; \text{or}\; 2<\mu <\frac{5}{2} \; \text{or}\; \mu >\frac{5}{2}$.
   
   \item $\mu= 1 \; \text{ if}\; w\neq 1$. 
   
   \item $\mu = 2\; \text{ if }\; 1<w\leq 2 \; \text{or}\; w<1$.
   
   \item $\mu= \frac{5}{2}\; \text{ if }\; w\geq \frac{4}{3}$.
   
   \item $\mu=\frac{5}{2}$ if $w= -2$.   
\end{itemize}

In this case, the~equation of state for radiation is not recovered.

For $\epsilon=1$, Equations \eqref{compatibility_1}
and \eqref{compatibility_2}  lead~to 
\begin{itemize}
    \item $w=\frac{-7+\sqrt{8 \mu  (2 \mu -9)+105}}{4 (\mu -1)} \; \text{ if }\; \mu <1 \; \text{or }\; 1<\mu <2 \; \text{or}\; 2<\mu <\frac{5}{2} \; \text{or}\; \mu >\frac{5}{2}$.  In this case, the~equation of state for radiation is recovered for $\mu=41/8$.
    \item $\mu =2 \; \text{ if} \; w\geq 2$.
    \item $\mu =\frac{5}{2}\;\text{ if }\; 1<w\leq \frac{4}{3} \; \text{or}\;    w<1$.
    \item $\mu =1$ if $w= -\frac{5}{7}$. 
    \item $\mu=2$ if $w= -\frac{1}{2}$.
\end{itemize}

The second approach consisted of solving the Riccati equation~\eqref{Rccati2} for $\mu\neq 1$ 
 following  substitution of $p=w \rho$ in \eqref{NewRaych} and removing $\rho$ using \eqref{NewFried}. The~exact solution is \eqref{exactRicatti2},
where $H_0$ is the current value of $H$ at $t=t_0$.  From~\eqref{NewRaych} and \eqref{NewCons}, we found
that compatibility conditions have to be satisfied for all $t$. It is necessary that $b_1=0$.  
Therefore, to~obtain solutions, we solved \eqref{compatibility_21}
and \eqref{compatibility_22} simultaneously for $\mu$ and $w$. We have the following~solutions: 
\begin{itemize}
    \item $w=-1 \; \text{ if }\; \mu \geq 3$.
    
    \item $w=\frac{7+ \sqrt{8 \mu  (2 \mu -9)+105}}{4(1-\mu)}\; \text{ if }\; 2<\mu <\frac{5}{2}\; \text{or}\; \mu >\frac{5}{2}\; \text{or}\; \mu <2$. 
    
    \item $w=\frac{-7 +\sqrt{8 \mu  (2 \mu -9)+105}}{4 (\mu -1)} \; \text{ if }\; 2<\mu <\frac{5}{2}\; \text{or}\; \frac{5}{2}<\mu \leq \frac{1}{4} \left(9+\sqrt{201}\right)\; \text{or}\; \mu <2$.
    
    \item $\mu=2 \; \text{ if }\; w\leq -\frac{4}{3}$.
    
    \item $\mu=\frac{5}{2}\; \text{ if }\; w\geq -\frac{8}{9}$.
    
    \item $\mu= 1$ if $w=  -\frac{5}{7}$. 
   
   \item $\mu= 2$ if $w= -\frac{1}{2}$. 
   
   \item $\mu= \frac{5}{2}$ if $w= -2$. 
   \end{itemize}
   
Upon physical consideration, we remove the cases with $w=-1$ and $\mu \in\{1,2\}$ and~we assume $-1<w<1$. 

Using the two previous approaches, one obtains power-law solutions of the type $a=(t/t_0)^{\alpha}$. Additionally, one has to impose two compatibility conditions which allow particular solutions to  $(\mu, w)$. That means that any solution of the power law type is indeed a particular exact solution of the system, but~not the general solution. To~obtain an exact solution that gives the general solution of the system (for any value of the free parameters that have an integration constant), we solved the Riccati Equation~\eqref{Riccati} independent of the EoS. This gave us solution \eqref{V}. This result is generic since it does not require specifying the EoS. That led to 
 $a(\tau)$ given by \eqref{(65)}, $H(\tau)$ given by~\eqref{(66)},
$p(\tau)$ given by \eqref{(68)}, $\rho(\tau)$ given by \eqref{(69)}, $q(\tau)$ given by \eqref{(67)},   
 $w_{\text{eff}}(\tau)$ given by \eqref{(70)} and $\Omega_{\text{m}}(\tau)$ given by 
\eqref{(71)}, where we defined the new time variable $\tau=H_{0}t$, such that $\alpha_0=H_{0}t_0$,  $ c = \frac{-2 \mu +r-6 \alpha_0+9}{2 \mu +r+6 \alpha_0-9}$,  and~ $r= \sqrt{8 \mu  (2 \mu -9)+105}$.

We have
\begin{equation}\label{Ominfty}
\lim_{t \rightarrow \infty}  w_{\text{eff}} (t) = \frac{-7+\sqrt{8 \mu  (2 \mu -9)+105}}{4 (\mu -1)}, \quad 
  \lim_{t \rightarrow \infty}   \Omega_{\text{m}}  (t) = \frac{5-\sqrt{8 \mu  (2 \mu -9)+105}}{2 (\mu -2)}, 
\end{equation}
such that $-1\leq  \lim_{t \rightarrow \infty}  w_{\text{eff}} (t) \leq -1/3$ for $\mu \leq \frac{5}{2}$ and~ 
$0\leq   \lim_{t \rightarrow \infty}   \Omega_{\text{m}}  (t) \leq 1$ for $1\leq \mu \leq \frac{5}{2}$.
 For large $t$, we have the asymptotic solution
\begin{equation}
H (t) \simeq  \frac{1}{6t} \left(9-2 \mu +\sqrt{8 \mu  (2 \mu -9)+105}\right),
\end{equation}
and \begin{equation}
\lim_{t \rightarrow \infty} \alpha (t) =    \frac{1}{6} \left(9-2 \mu +\sqrt{8 \mu  (2 \mu -9)+105}\right)\geq 0,
\end{equation}
where $\alpha(t)=t H$ is the age~parameter. 

Finally, combining the solution of Bernoulli’s Equation \eqref{eqm} and~inequality \eqref{ineq} and then solving the differential inequality \eqref{eqN} and~approximating the different quadrature, we have
\begin{align*}
& E(z)  \simeq \frac{1}{\alpha_0} \left[ \frac{1}{9} \left((9-2 \mu)\left(1-  (1+z)^{-3}\right) + 3 m_0^{\frac{1}{2}}\right)^2 (1+z)^{6} + A^2 \right]^{\frac{1}{2}} \left(1+z\right)^\frac{3}{\sqrt{9 A^2+(9-2 \mu )^2}}
\end{align*}
and 
\begin{equation*}
t(z) \simeq t_{0} (1+z)^{-\frac{3}{\sqrt{9 A^2+(9-2 \mu )^2}}},
\end{equation*}
as $z\rightarrow -1$, 
where $\alpha_0$ is defined by \eqref{alpha_0},
$A$ is defined by \eqref{(EQ140)}, satisfying  inequality \eqref{inequality} and~$m_0= \alpha_0^2-A^2$. This is an accurate approximation of $E(z)$ as $z\rightarrow -1$ provided  $2\leq \mu \leq \frac{1}{10} \left(63+\sqrt{849}\right)\lesssim 9.21376$.  Moreover, $A \in [0, A_+]$ for  $2\leq\mu \leq3.25162$ or~$7.59791\leq\mu \leq9.21376$.

%%%%%%%%%%%%%%%%%%%%%%%%%%%%%%%%%%
\section{Cosmological~Constraints}
\label{Sect:VII}
%%%%%%%%%%%%%%%%%%%%%%%%%%%%%%%%%
In this section, to~study the capability of the models obtained in fractional cosmology to describe the late-time accelerated Universe expansion, we shall constrain the free parameters with the SNe Ia data and OHD. In~particular, for~the first one, we consider the Pantheon sample~\cite{Scolnic_Complete_2018}, which consists of $1048$ supernovae data points in the redshift range $0.01\leq z\leq 2.3$. On~the other hand, we consider the OHD compiled by Magaña~et~al. \cite{Magana_Cardassian_2018}, which consists of $51$ data points in the redshift range $0.07\leq z\leq 2.36$. 

For the constraints, we compute the best-fit parameters and their respective confidence regions at $1\sigma(68.3\%)$, $2\sigma(95.5\%)$ and $3\sigma(99.7\%)$  confidence levels (CLs) with the affine-invariant Markov chain Monte Carlo (MCMC) method~\cite{Goodman_Ensemble_2010}, implemented in the pure-Python code emcee \cite{Foreman_emcee_2013} by~setting $35$ chains or ``walkers''. As~a convergence test, we computed the~autocorrelation time $\tau_{corr}$ of the chains provided by the emcee module at~every $50$th step. Hence, if~the current step is larger than $50\tau_{corr}$ and the values of $\tau_{corr}$ changed by less than $1\%$, then we will consider that the chains are converged and~the constraint is stopped. The first $5\tau_{corr}$ steps are thus discarded as ``burn-in'' steps. This convergence test was complemented with the calculation of the mean acceptance fraction, which must have a value between $0.2$ and $0.5$ \cite{Foreman_emcee_2013} and can be modified by the stretch move provided by the emcee module.

For this Bayesian statistical analysis, we need to construct the following Gaussian likelihood:
\begin{equation}\label{likelihood}
    \mathcal{L}=\mathcal{N}\exp{\left(-\frac{\chi_{I}^{2}}{2}\right)},
\end{equation}
where $\mathcal{N}$ is a normalization constant, which does not influence the MCMC analysis, and~$\chi_{I}^{2}$ is the merit function of each dataset considered, i.e.,~$I$ stands for SNe Ia, OHD and~their joint analysis. In~the following subsections, we will briefly describe the construction of the merit function of each dataset, and~we will present the main results and~discussions.

%%%%%%%%%%%%%%%%%%%%%%%%%%%%%%%%%%%%%%%%%%%%%%%%
\subsection{Observational Hubble Parameter~Data}
\label{Sect:VII.1}
%%%%%%%%%%%%%%%%%%%%%%%%%%%%%%%%%%%%%%%%%%%%%%%%
The merit function for the OHD is constructed as
\begin{equation}\label{meritOHD}
    \chi_{OHD}^{2}=\sum_{i=1}^{51}{\left[\frac{H_{i}-H_{th}(z_{i},\theta)}{\sigma_{H,i}}\right]^{2}},
\end{equation}
where $H_{i}$ is the observational Hubble parameter at redshift $z_{i}$ with an associated error $\sigma_{H, i}$, all of them provided by the OHD sample, $H_{th}$ is the theoretical Hubble parameter at the same redshift and~$\theta$ encompasses the free parameters of the model under study. It is important to mention that the current value of the Hubble parameter, $H_{0}$, is a free parameter of the model, which for the constraint is written as $H_{0}=100\frac{\text{km/s}}{\text{Mpc}}h$, where~$h$ is dimensionless. Considering that we expect a value of $H_{0}$ between the value obtained from Planck CMB of $H_{0}=67.4$ \cite{Planck:2018vyg} for the $\Lambda$CDM model and the value obtained by A. G. Riess~et~al. of $H_{0}=74.03$ \cite{Riess:2019cxk} in a model-independent way, then we consider for $h$ the flat prior $h\in F(0.4, 1)$.

%%%%%%%%%%%%%%%%%%%%%%%%%%%%%%%%%%%%
\subsection{Type Ia Supernovae~Data}
\label{Sect:VII.2}
%%%%%%%%%%%%%%%%%%%%%%%%%%%%%%%%%%%%
Similarly to the OHD, the~merit function for the SNe Ia data is constructed as
\begin{equation}\label{meritSNeIa}
    \chi_{SNe}^{2}=\sum_{i=1}^{1048}{\left[\frac{\mu_{i}-\mu_{th}(z_{i},\theta)}{\sigma_{\mu,i}}\right]^{2}},
\end{equation}
where $\mu_{i}$ is the observational distance modulus of each SNe Ia at redshift $z_{i}$ with an associated error $\sigma_{\mu, i}$, $\mu_{th}$ is the theoretical distance modulus for each SNe Ia at the same redshift and~$\theta$ encompasses the free parameters of the model under study. Following this line, for~a spatially flat FLRW spacetime, the~theoretical distance modulus is given by
\begin{equation}\label{theoreticaldistance}
    \mu_{th}(z_{i},\theta)=5\log_{10}{\left[\frac{d_{L}(z_{i},\theta)}{\text{Mpc}}\right]}+\Bar{\mu},
\end{equation}
where $\Bar{\mu}=5\left[\log_{10}{\left(c\right)}+5\right]$ and where~$c$ is the speed of light given in units of $\text{km/s}$. The~above expression relates the merit function with the theoretical Hubble parameter through the luminosity distance, $d_{L}$, as follows
\begin{equation}\label{luminosity}
    d_{L}(z_{i},\theta)=(1+z_{i})\int_{0}^{z_{i}}{\frac{dz'}{H_{th}(z',\theta)}}.
\end{equation}

On the other hand, the~distance estimator used in the Pantheon sample is obtained by a modified version of Tripp's formula~\cite{Tripp_Luminosity_1998}, with~two of the three nuisance parameters calibrated to zero with the BEAMS with bias correction (BBC) method~\cite{Kessler_Correcting_2017}. Hence, the~observational distance modulus for each SNe Ia reads
\begin{equation}\label{pantheondistance}
    \mu_{i}=m_{B,i}-\mathcal{M},
\end{equation}
where $m_{B, i}$ is the corrected apparent B-band magnitude of a fiducial SNe Ia at redshift $z_{i}$ with an associated error $\sigma_{m_B, i}$, all of them provided by the Pantheon sample (currently available online in the GitHub repository \url{https://github.com/dscolnic/Pantheon}  (accessed on 28 April 2023).
The~corrected apparent B-band magnitude $m_{B, i}$ for each SNe Ia with their respective redshifts $z_{i}$ and errors $\sigma_{m_{B}, i}$ are available in the document \textit{lcparam\_full\_long.txt}), and~$\mathcal{M}$ is a nuisance parameter which must be jointly estimated with the free parameters $\theta$ of the theoretical model. Furthermore, the~Pantheon sample provides the systematic uncertainties in the BBC approach, $\mathbf{C}_{sys}$ (currently available online in the GitHub repository \url{https://github.com/dscolnic/Pantheon}  (accessed on 28 April 2023) in the document \textit{sys\_full\_long.txt}). Therefore, we can rewrite the merit function \eqref{meritSNeIa} in matrix notation denoted by bold  symbols as 
\begin{equation}\label{meritSNeIamatrix}
    \chi_{SNe}^{2}=\mathbf{M}(z,\theta,\mathcal{M})^{\dagger}\mathbf{C}^{-1}\mathbf{M}(z,\theta,\mathcal{M}),
\end{equation}
where $[\mathbf{M}(z,\theta,\mathcal{M})]_{i}=m_{B,i}-\mu_{th}(z_{i},\theta)-\mathcal{M}$ and $\mathbf{C}=\mathbf{D}_{stat}+\mathbf{C}_{sys}$ is the total uncertainty covariance matrix, where $\mathbf{D}_{stat}=diag(\sigma_{m_{B},i}^{2})$ is the statistical uncertainty of $m_{B}$.

Finally, one can marginalize over the nuisance parameters $\Bar{\mu}$ and $\mathcal{M}$ by~defining $\Bar{\mathcal{M}}=\Bar{\mu}+\mathcal{M}$. Then, the~merit function \eqref{meritSNeIamatrix} can be expanded as~\cite{Lazkos_Exploring_2005}
\begin{equation}\label{meritSNeIaexpanded}
    \chi_{SNe}^{2}=A(z,\theta)-2B(z,\theta)\Bar{\mathcal{M}}+C\Bar{\mathcal{M}}^{2},
\end{equation}
where
\begin{equation}\label{defofA}
    A(z,\theta)=\mathbf{M}(z,\theta,\Bar{\mathcal{M}}=0)^{\dagger}\mathbf{C}^{-1}\mathbf{M}(z,\theta,\Bar{\mathcal{M}}=0),
\end{equation}
\begin{equation}\label{defofB}
    B(z,\theta)=\mathbf{M}(z,\theta,\Bar{\mathcal{M}}=0)^{\dagger}\mathbf{C}^{-1}\mathbf{1},
\end{equation}
\begin{equation}
    C=\mathbf{1}\mathbf{C}^{-1}\mathbf{1}.
\end{equation}

Therefore, by~minimizing the expanded merit function \eqref{meritSNeIaexpanded} with respect to $\Bar{\mathcal{M}}$,  $\Bar{\mathcal{M}}=B(z,\theta)/C$ is obtained and~the expanded merit is function reduced to
\begin{equation}\label{meritSNeIaminimize}
    \chi_{SNe}^{2}=A(z,\theta)-\frac{B(z,\theta)^{2}}{C},
\end{equation}
which depends only on the free parameters of the theoretical~model.

It is essential to mention that the expanded and minimized merit function \eqref{meritSNeIaminimize} provides the same information as the merit function \eqref{meritSNeIamatrix}. This is a consequence of the fact that the best-fit parameters minimize the merit function. Therefore, the~evaluation of the best-fit parameters in the merit function can be used as an indicator of the goodness of the fit independently of the dataset used; the smaller the value of $\chi_{min}^{2}$, the~better the~fit. 

%%%%%%%%%%%%%%%%%%%%%%%%%%%%%%%%%%%%%%%%%%%%%%%%%%%%%%%%%%%%%%%%%%%%%%%%
\subsection{Joint Analysis and Theoretical Hubble Parameter~Integration}
\label{Sect:VII.3}
%%%%%%%%%%%%%%%%%%%%%%%%%%%%%%%%%%%%%%%%%%%%%%%%%%%%%%%%%%%%%%%%%%%%%%%%
The merit function for the joint analysis is constructed directly as
\begin{equation}\label{meritjoint}
    \chi_{joint}^{2}=\chi_{OHD}^{2}+\chi_{SNe}^{2},
\end{equation}
with $\chi^{2}_{OHD}$ and $\chi^{2}_{SNe}$ given by Equations \eqref{meritOHD} and \eqref{meritSNeIaminimize}, respectively. Following this line, note how in the merit function of the two datasets, the~respective model is considered through the (theoretical) Hubble parameter as a function of the redshift (see Equations~\eqref{meritOHD} and \eqref{luminosity}). Hence, for~the constraint, we numerically integrate the system given by Equations \eqref{eq(134)} and \eqref{eq(135)}, which represents a system for the variables $(\alpha,t)$ as a function of $s=-\ln{\left(1+z\right)}$, and~for which we consider the initial conditions $\alpha(s=0)\equiv\alpha_{0}=t_{0}H_{0}$ and $t(s=0)\equiv t_{0}=\alpha_{0}/H_{0}$. Then, the~Hubble parameter is obtained numerically by $H_{th}(z)=\alpha(z)/t(z)$. For~this integration, we consider the NumbaLSODA code, a~python wrapper of the LSODA method in ODEPACK to C+ {(currently available online in the GitHub repository \url{https://github.com/Nicholaswogan/numbalsoda}  (accessed on 28 April 2023})). Furthermore, for~further comparison, we also constrain the free parameters of the $\Lambda$CDM model, whose respective Hubble parameter as a function of the redshift is given by
\begin{equation}\label{HLCDM}
    H(z)=H_{0}\sqrt{\Omega_{m,0}(1+z)^{3}+1-\Omega_{m,0}}.
\end{equation}

Finally, based on the analysis made in Section~\ref{Sect:V.1}, we consider the parameterization given by Equation \eqref{alpha_0} for the free parameter $\alpha_{0}$. Therefore, the~free parameters of the fractional cosmological model are $\theta=\{h,\mu,\epsilon_{0}\}$ and~the free parameters of the $\Lambda$CDM model are $\theta=\{h,\Omega_{m,0}\}$. For~the free parameters $\mu$, $\epsilon_{0}$ and~$\Omega_{m,0}$, we consider the following flat priors: $\mu\in F(1,4)$, $\epsilon_{0}\in F(-0.1,0.1)$ and~$\Omega_{m,0}\in F(0,1)$. It is important to mention that due to a degeneracy between $H_{0}$ and $\mathcal{M}$, the~SNe Ia data are not able to constrain the free parameter $h$ (as a reminder, $H_{0}=100\frac{\text{km/s}}{\text{Mpc}}h$), contrary to the case for the OHD and, consequently, in~the joint analysis. 
Thus, the~posterior distribution of $h$ for the SNe Ia data is expected to cover all the prior distributions. On~the other hand, the~prior is chosen as $\epsilon_{0}$  because $\epsilon_0$ is a measure of the limiting value of the relative error in the age parameter $t H$ when it is approximated by $t_0 H_0$ as given by Equation \eqref{alpha_0}. For~the mean value $\epsilon_0=0$, we acquire $\alpha_0=\frac{1}{6} (-2 \mu +r+9)$, which implies $c=0$. Then, we have  the leading term  for $E(z)$ defined by \eqref{(142)}. The~lower prior of $\mu$ is because the Hubble parameter \eqref{3} becomes negative when $\mu<1$ in~the absence of matter, as~we can see from Section~\ref{Sect:III}.

%%%%%%%%%%%%%%%%%%%%%%%%%%%%%%%%%%%
\subsection{Results and~Discussion}
\label{Sect:VII.4}
%%%%%%%%%%%%%%%%%%%%%%%%%%%%%%%%%%%
In Table~\ref{tab:emcee}, we present the total steps, the~mean acceptance fraction and~the autocorrelation time, $\tau_{corr}$, of each free parameter obtained when the convergence test is fulfilled during our MCMC analysis for both the~fractional cosmological model and the $\Lambda$CDM model. The~values of the mean acceptance fraction are obtained for a value of the stretch move of $a=7$ for the $\Lambda$CDM model and $a=3.5$ for the fractional cosmological~model.

\begin{table}[H]
 %   \centering
    \caption{The total number of steps, means acceptance fraction (MAF) and~autocorrelation time, $\tau_{corr}$, for the free parameters of the fractional cosmological model and the $\Lambda$CDM model. These values are obtained when the convergence test described in Section~\ref{Sect:VII} is fulfilled for an MCMC analysis with $35$ chains, a~value of the stretch move of $a=7$ for the $\Lambda$CDM model and $a=3.5$ for the fractional cosmological model and~for the flat priors $h\in F(0.4,1)$, $\Omega_{m,0}\in F(0,1)$, $\mu\in F(1,4)$ and $\epsilon_{0}\in F(-0.1,0.1)$.
      \label{tab:emcee}}
\newcolumntype{C}{>{\centering\arraybackslash}X}
\begin{tabularx}{\textwidth}{ccCCCCC}
    \toprule
         & & & \multicolumn{4}{c}{\boldmath{$\tau_{corr}$}} \\
        \cmidrule{4-7}
       \textbf{Data} & \textbf{Total Steps} & \textbf{MAF} & \boldmath{$h$} & \boldmath{$\Omega_{m,0}$} & \boldmath{$\mu$} & \boldmath{$\epsilon_{0}$}\\
        \midrule
        \multicolumn{7}{c}{\boldmath{$\Lambda$}\textbf{CDM model}} \\ 
         SNe Ia & $1300$ & $0.328$ & $24.0$ & $20.8$ & $\cdots$ & $\cdots$ \\
         OHD & $950$ & $0.364$ & $16.1$ & $15.4$ & $\cdots$ & $\cdots$ \\
         SNe Ia + OHD  & $850$ & $0.364$ & $15.4$ & $16.5$ & $\cdots$ & $\cdots$ \\
        \midrule
        \multicolumn{7}{c}{\textbf{Fractional cosmological model}} \\
         SNe Ia  & $5250$ & $0.339$ & $50.9$ & $\cdots$ & $104.3$ & $69.9$ \\
         OHD  & $3150$ & $0.377$ & $35.9$ & $\cdots$ & $62.3$ & $58.4$ \\
         SNe Ia + OHD  & $1900$ & $0.413$ & $27.8$ & $\cdots$ & $33.8$ & $32.1$ \\
        \bottomrule
    \end{tabularx}
\end{table}

\begin{figure}[H]
  %  \centering
    \includegraphics[scale = 0.40]{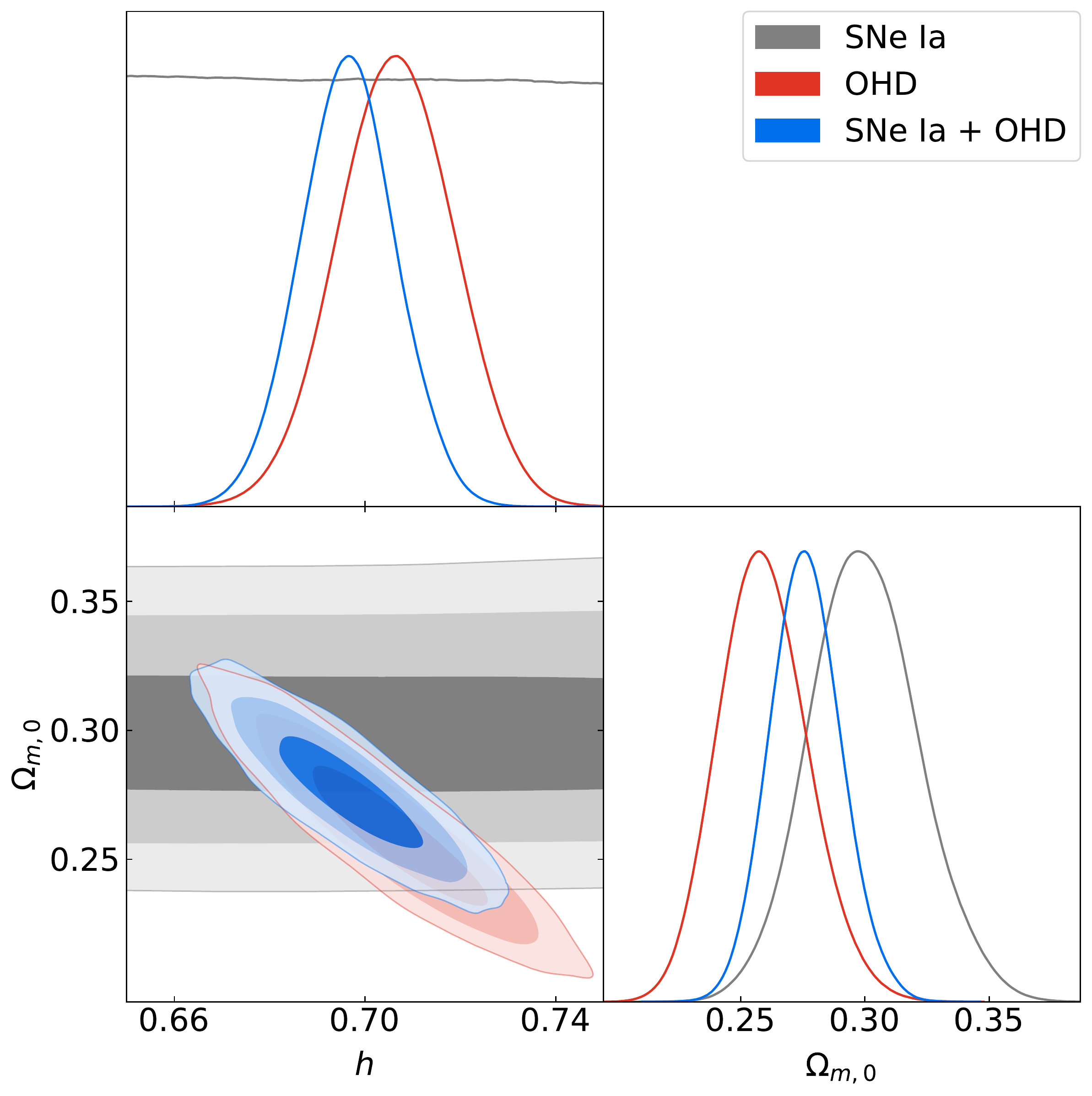}
    \caption{Posterior distribution and joint admissible regions of the free parameters $h$ and $\Omega_{m,0}$ for the $\Lambda$CDM model, obtained by the MCMC analysis described in Section~\ref{Sect:VII}. The~admissible joint regions correspond to $1\sigma(68.3\%)$, $2\sigma(95.5\%)$ and~$3\sigma(99.7\%)$ of confidence level (CL), respectively. The~best-fit values for each model free parameter are shown in Table~\ref{tab:bestfits}.}
    \label{fig:TriangleLCDM}
\end{figure}

\begin{figure}[H]
    %\centering
    \includegraphics[scale = 0.40]{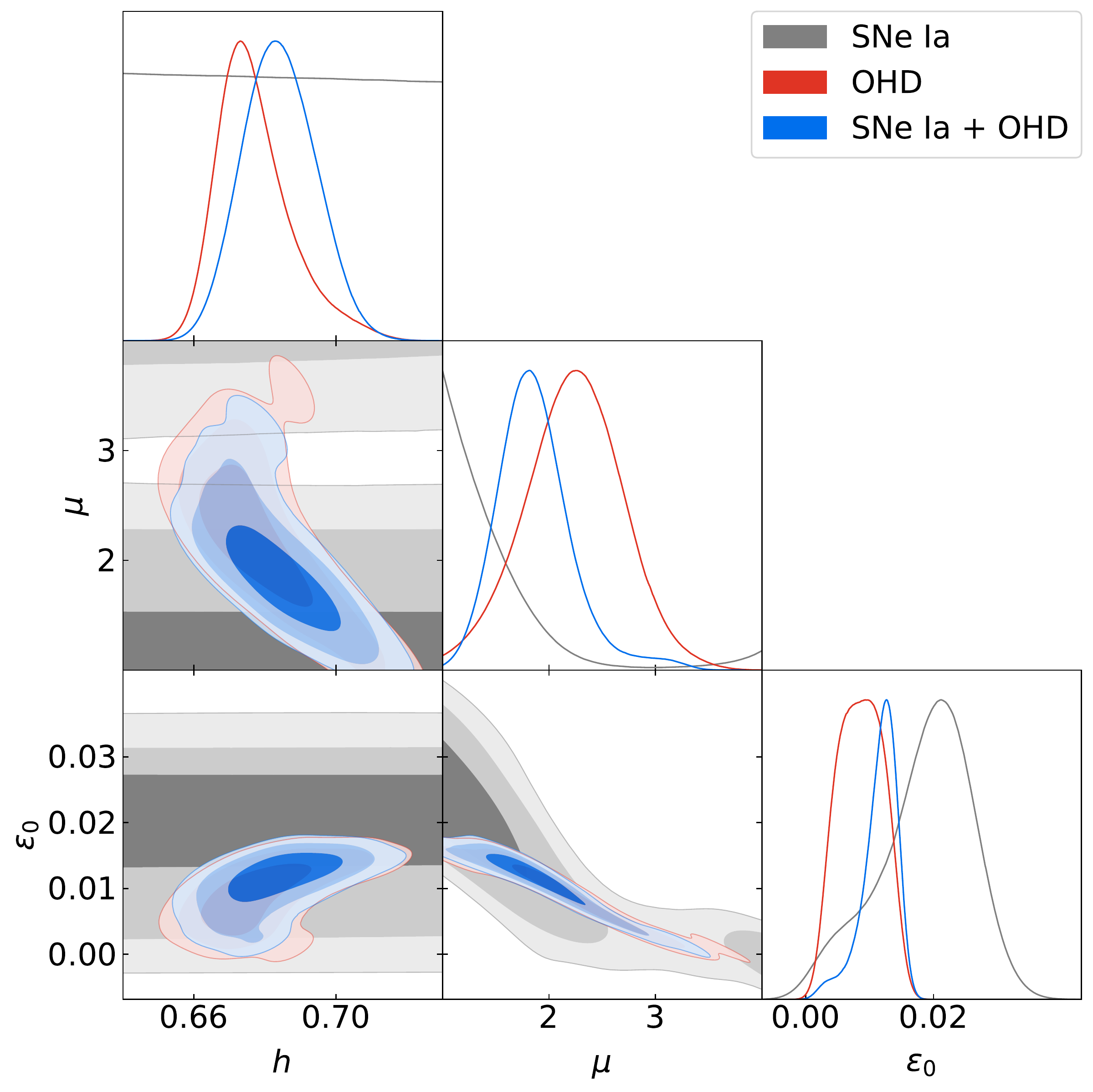}
    \caption{Posterior distribution and joint admissible regions of the free parameters $h$, $\mu$ and~$\epsilon_{0}$ for the fractional cosmological model, obtained by the MCMC analysis described in Section \ref{Sect:VII}. The~admissible joint regions correspond to $1\sigma(68.3\%)$, $2\sigma(95.5\%)$ and~$3\sigma(99.7\%)$  confidence levels (CLs), respectively. The~best-fit values for each model free parameter are shown in Table~\ref{tab:bestfits}.}
    \label{fig:TriangleFractional}
\end{figure}

The best-fit values of the free parameters space for the $\Lambda$CDM model and the fractional cosmological model, obtained for the SNe Ia data, OHD and~in their joint analysis, with~their corresponding $\chi^{2}_{min}$ criteria, are presented in Table~\ref{tab:bestfits}. The~uncertainties correspond to $1\sigma$, $2\sigma$ and~$3\sigma$ CL. In~Figures~\ref{fig:TriangleLCDM} and \ref{fig:TriangleFractional}, we depict the posterior distribution and joint admissible regions of the free parameter space of the $\Lambda$CDM model and the fractional cosmological model, respectively. The~joint admissible regions correspond to $1\sigma$, $2\sigma$ and~$3\sigma$ CL. Due to the degeneracy between $H_{0}$ and $\mathcal{M}$, the~distribution of $h$ for the SNe Ia data was not represented in its full parameter~space.

\begin{table}[H]
    \caption{Best-fit values and $\chi^{2}_{min}$ criteria for the fractional cosmological model with free parameters $h$, $\mu$ and~$\epsilon_{0}$ and for the $\Lambda$CDM model with free parameters $h$ and $\Omega_{m,0}$. The~values were obtained by the MCMC analysis described in Section~\ref{Sect:VII} for the SNe Ia data, OHD and~their joint analysis. The~uncertainties presented correspond to $1\sigma(68.3\%)$, $2\sigma(95.5\%)$ and~$3\sigma(99.7\%)$  confidence levels (CLs), respectively. The~$\Lambda$CDM model is used as a reference~model.}
    \label{tab:bestfits}
      \setlength{\tabcolsep}{4mm}
      \resizebox{\textwidth}{!}{
    \begin{tabular}{cccccc}
        \toprule
         & \multicolumn{4}{c}{\textbf{Best-Fit Values}} & \\ 
         \cmidrule{2-6}
  
        \textbf{Data} & \boldmath{$h$} & \boldmath{$\Omega_{m,0}$} & \boldmath{$\mu$} & \boldmath{$\epsilon_{0}\times 10^{2}$} & \boldmath{$\chi_{min}^{2}$} \\
        \midrule
        \multicolumn{6}{c}{\boldmath{$\Lambda$}\textbf{CDM model}} \\ %[1ex]%\\
         SNe Ia & $0.692_{-0.120\;-0.278\;-0.292}^{+0.209\;+0.296\;+0.307}$ & $0.299_{-0.021\;-0.042\;-0.059}^{+0.022\;+0.046\;+0.068}$ & $\cdots$ & $\cdots$ & $1026.9$ \\[1.5ex]% \\
         OHD & $0.706_{-0.012\;-0.024\;-0.036}^{+0.012\;+0.024\;+0.035}$ & $0.259_{-0.017\;-0.033\;-0.047}^{+0.018\;+0.038\;+0.059}$ & $\cdots$ & $\cdots$ & $27.5$ \\ [1.5ex]%\\
         SNe Ia + OHD & $0.696_{-0.010\;-0.020\;-0.029}^{+0.010\;+0.020\;+0.029}$ & $0.276_{-0.014\;-0.027\;-0.040}^{+0.014\;+0.030\;+0.043}$ & $\cdots$ & $\cdots$ & $1056.3$ \\%[1.5ex]% \\
        \midrule
        \multicolumn{6}{c}{\textbf{Fractional cosmological model}} \\ [1ex]%\\
         SNe Ia & $0.696_{-0.204\;-0.284\;-0.295}^{+0.215\;+0.293\;+0.302}$ & $\cdots$ & $1.340_{-0.245\;-0.328\;-0.339}^{+0.492\;+2.447\;+2.651}$ & $1.976_{-0.905\;-1.848\;-2.067}^{+0.599\;+1.133\;+1.709}$ & $1028.1$ \\ [1.5ex]%\\
         OHD  & $0.675_{-0.008\;-0.015\;-0.021}^{+0.013\;+0.029\;+0.041}$ & $\cdots$ & $2.239_{-0.457\;-0.960\;-1.190}^{+0.449\;+0.908\;+1.386}$ & $0.865_{-0.407\;-0.657\;-0.773}^{+0.395\;+0.650\;+0.793}$ & $29.7$ \\ [1.5ex]%\\
         SNe Ia + OHD  & $0.684_{-0.010\;-0.020\;-0.027}^{+0.011\;+0.021\;+0.031}$ & $\cdots$ & $1.840_{-0.298\;-0.586\;-0.773}^{+0.343\;+1.030\;+1.446}$ & $1.213_{-0.310\;-0.880\;-1.057}^{+0.216\;+0.383\;+0.482}$ & $1061.1$ \\%[1.5ex]% \\
   \bottomrule
  \end{tabular}}
\end{table}

\unskip

\begin{figure}[H]
  %  \centering
    \includegraphics[scale = 0.65]{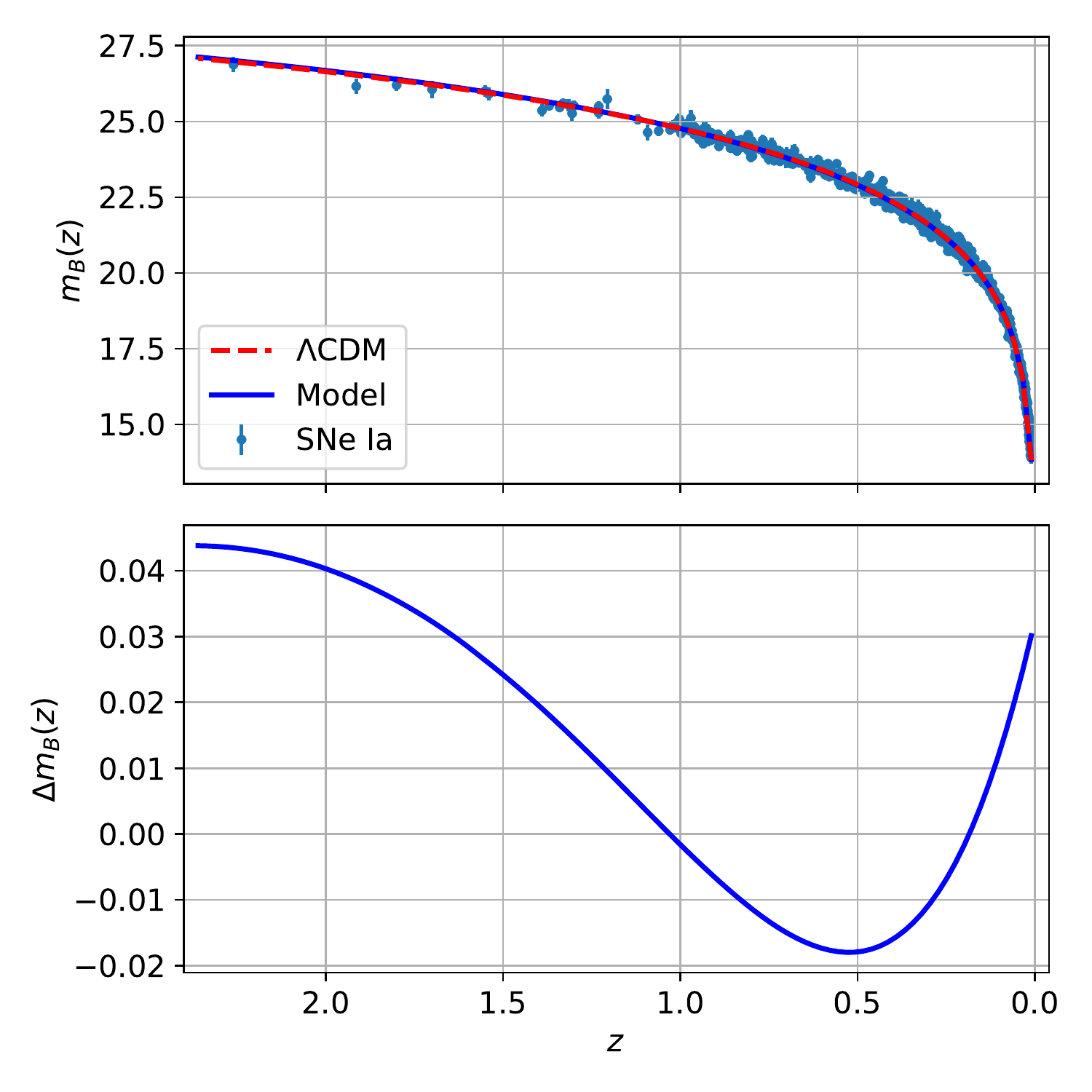}
    \caption{(\textbf{Top panel}) Theoretical apparent B-band magnitude for the $\Lambda$CDM model (red dashed line) and the fractional cosmological model (solid blue line) as a function of the redshift $z$, contrasted with the pantheon dataset. (\textbf{Bottom panel}) Variation in the theoretical apparent B-band magnitude of the fractional cosmological model compared to the $\Lambda$CDM model as a function of the redshift $z$. The~curve is obtained through the expression $\Delta m_{B}=m_{B, Model}-m_{B,\Lambda CDM}$. The~figures were obtained using the best-fit values for the SNe Ia+OHD data presented in Table~\ref{tab:bestfits}.}
    \label{fig:SNeFractional}
\end{figure}

\begin{figure}[H]
 %   \centering
    \includegraphics[scale = 0.65]{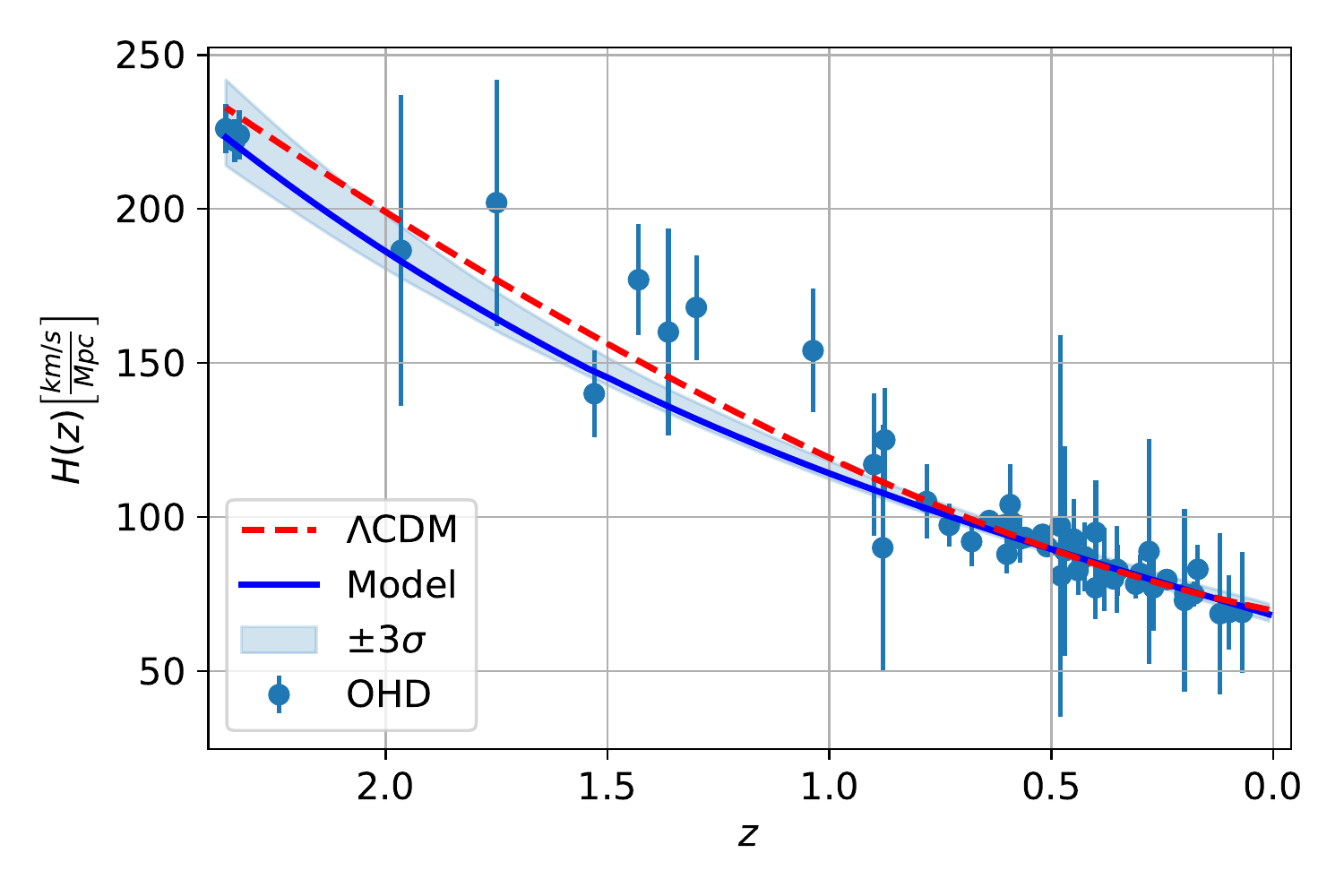}
    \caption{Theoretical Hubble parameter for the $\Lambda$CDM model (red dashed line) and the fractional cosmological model (solid blue line) as a function of the redshift $z$, contrasted with the OHD sample. The~shaded curve represents the confidence region of the Hubble parameter for the fractional cosmological model at a $3\sigma(99.7\%)$  confidence level (CL). The~figure was obtained using the best-fit values for the SNe Ia+OHD data presented in Table~\ref{tab:bestfits}.}
    \label{fig:OHDFractional}
\end{figure}

From the values for the $\chi^{2}_{min}$ criteria presented in Table~\ref{tab:bestfits}, it is possible to see that the $\Lambda$CDM model is the best model to constrain the SNe Ia data, OHD and~SNe Ia + OHD data. Nevertheless, the~fractional cosmological model studied in this paper exhibits values of the $\chi^{2}_{min}$ criteria close to the values of the $\Lambda$CDM model, with~differences of $1.2$ for the SNe Ia data, $2.2$ for the OHD data and~$4.8$ for their joint analysis. Thus, this fractional cosmological model is suitable for describing the SNe Ia and OHD data, as~can be seen from Figures~\ref{fig:SNeFractional} and~\ref{fig:OHDFractional}, which are characterized by accounting for a universe that experiences a transition between a deceleration expansion phase and an accelerated one. Therefore, fractional cosmology can be considered an alternative valid cosmological model to describe the late-time Universe. It is essential to mention that the core of this work is to probe this possibility by studying a particular model; the $\Lambda$CDM model is used only as a reference model for this~aim.

The analysis of the SNe Ia data leads to $h=0.696_{-0.295}^{+0.302}$, $\mu=1.340_{-0.339}^{+2.651}$ and  $\epsilon_0=\left(1.976_{-2.067}^{+1.709}\right)\times 10^{-2}$, which are the best-fit values at $3\sigma$ CL. In~this case, the~value obtained for $h$ cannot be considered as a best fit due to the degeneracy between $H_{0}$ and $\mathcal{M}$. On~the other hand, the~lower limit of the best fit for $\mu$ is very close to $1$. That is because the posterior distribution for this parameter is close to this value, as~seen from Figure~\ref{fig:TriangleFractional}. This indicates that a value of the SNe Ia data prefers $\mu<1$, but, as~a reminder, this value leads to a negative Hubble parameter in the absence of matter. However, as~can be seen from the same Figure~\ref{fig:TriangleFractional}, the~posterior distribution for these parameters is multi-modal (this explains the large value of $\tau_{corr}$ presented in Table~\ref{tab:emcee}) and, therefore, it is possible to obtain a best-fit value that satisfies $\mu>1$. It is important to mention that the OHD and the joint analysis do not experience this issue, which allows us to maintain the validity of the prior used for $\mu$.

\vspace{-6pt}

On the other hand, the~analysis from OHD leads to  $h=0.675_{-0.021}^{+0.041}$, $\mu=2.239_{-1.190}^{+1.386}$ and $\epsilon_0=\left(0.865_{-0.773}^{+0.793}\right)\times 10^{-2}$, which are the best-fit values at $3\sigma$ CL. In~this case, note how the OHD can properly constrain the free parameters $h$, $\mu$ and $\epsilon_{0}$, i.e.,~we obtain the best fit for the priors considered in our MCMC analysis. Furthermore, note how the posterior distribution of $\mu$ includes the value of $1$, as~seen from Figure~\ref{fig:TriangleFractional}, but~for a CL  greater than $3\sigma$.

Finally, the~joint analysis with data from SNe Ia + OHD leads to $h=0.684_{-0.027}^{+0.031}$, $\mu=1.840_{-0.773}^{+1.446}$ and $\epsilon_0=\left(1.213_{-1.057}^{+0.482}\right)\times 10^{-2}$, which are the best-fit values at $3\sigma$ CL. Focusing our analysis on these results, we can conclude that the region in which $\mu>2$ is not ruled out by observations. On~the other hand, these best-fit values lead to an age of the Universe with a value of $t_0=\alpha_0/H_0=25.62_{-4.46}^{+6.89}\;\text{Gyrs}$ at $3\sigma$ CL. Universe age is roughly double the one of the $\Lambda$CDM models, and is also in disagreement with the value obtained with globular clusters, with~a value of $t_0=13.5^{+0.16}_{-0.14}\pm 0.23$ \citep{Valcin:2021}. This discrepancy is a distinction of  fractional cosmology. This result also agrees with the analysis made in Section~8 of~\cite{Garcia-Aspeitia:2022uxz}, where the best-fit $\mu$-value was obtained from the reconstruction of $H(z)$ for different priors of $\mu$. The~results are summarized in Table~\ref{tab:BEST-FIT}. In \cite{Garcia-Aspeitia:2022uxz}, a set of 31 points obtained by differential age tools was considered, namely cosmic chronometers (CC), to represent the measurements of the Hubble parameter, which is cosmologically independent~\cite{Moresco:2016mzx} (in the present research we consider the datasets from~\cite{Magana_Cardassian_2018}, which consists of $51$ data points in the redshift range $0.07\leq z\leq 2.36$, 20 more points as compared with~\cite{Moresco:2016mzx}). The~1048 luminosity modulus measurements, known as the Pantheon sample, from~Type Ia Supernovae cover the region $0.01<z<2.3$  \cite{Scolnic_Complete_2018}. In~\cite{Garcia-Aspeitia:2022uxz}, it is unclear if the different priors used for $\mu$ lead to properly constraining $\mu$. Their analysis is inconclusive because of their present different values of $\mu$ for the different priors~used.

\begin{table}[H]
 \caption{The best-fit values $(\mu, t_0 )$ for different priors of $\mu$ derived in~\cite{Garcia-Aspeitia:2022uxz}.  }
    \label{tab:BEST-FIT}
\newcolumntype{C}{>{\centering\arraybackslash}X}
\begin{tabularx}{\textwidth}{CCC}
\toprule
 \textbf{Prior}    &  \boldmath{$\mu$} & \boldmath{$t_0$}\\\midrule
 $0< \mu<1$  & $0.50$ & $41.30\; \text{Gyrs}$ \\
 $ 1 < \mu < 3$ & $ 1.71 $ & $27.89\; \text{Gyrs}$ \\
$ 0 < \mu< 3$ & $1.15$ &$33.66\; \text{Gyrs}$\\\bottomrule
\end{tabularx}
\end{table}

In order to establish that this fractional cosmological model can describe a universe that experiences a transition from a decelerated expansion phase to an accelerated one, we computed the deceleration parameter $q=-1-\dot{H}/H^{2}$,  using the Riccati equation~\eqref{Riccati}, which leads to
\begin{equation}
q(\alpha(s))= 2 + \frac{2 (\mu -4) }{\alpha(s)}-\frac{(\mu -2) (\mu -1)}{\alpha^2(s)}.\label{DecelerationFinal}
\end{equation}

\begin{figure}[H]
   % \centering
    \includegraphics[scale = 0.65]{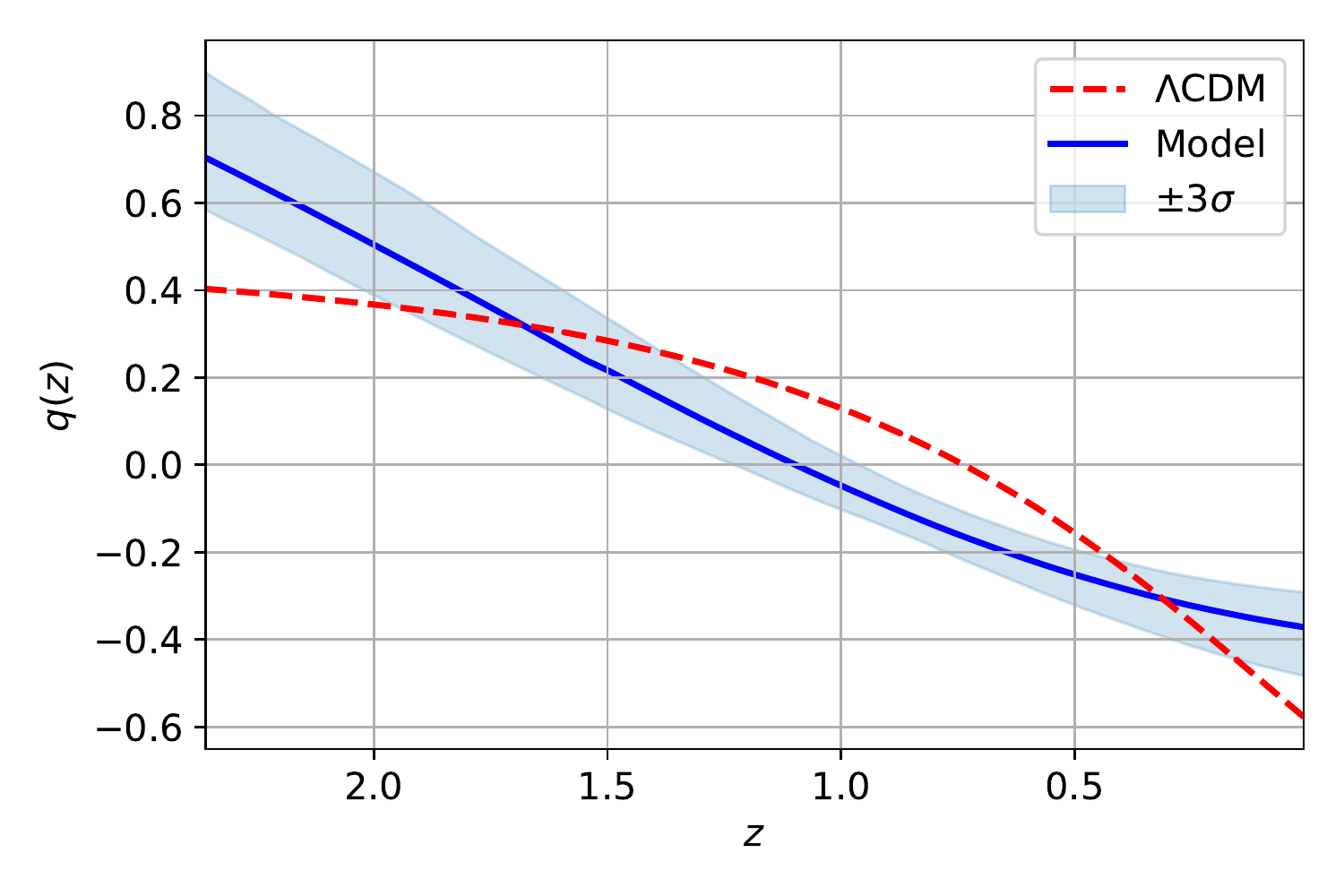}
    \caption{Deceleration parameters for the $\Lambda$CDM model (red dashed line) and the fractional cosmological model (solid blue line) as a function of the redshift $z$. The~shaded curve represents the confidence region of the deceleration parameter for the fractional cosmological model at a $3\sigma(99.7\%)$ confidence level (CL). The~figure was obtained using the best-fit values for the SNe Ia+OHD data presented in Table~\ref{tab:bestfits}.}
    \label{fig:qFractional}
\end{figure}

\begin{figure}[H]
 %   \centering
    \includegraphics[scale = 0.65]{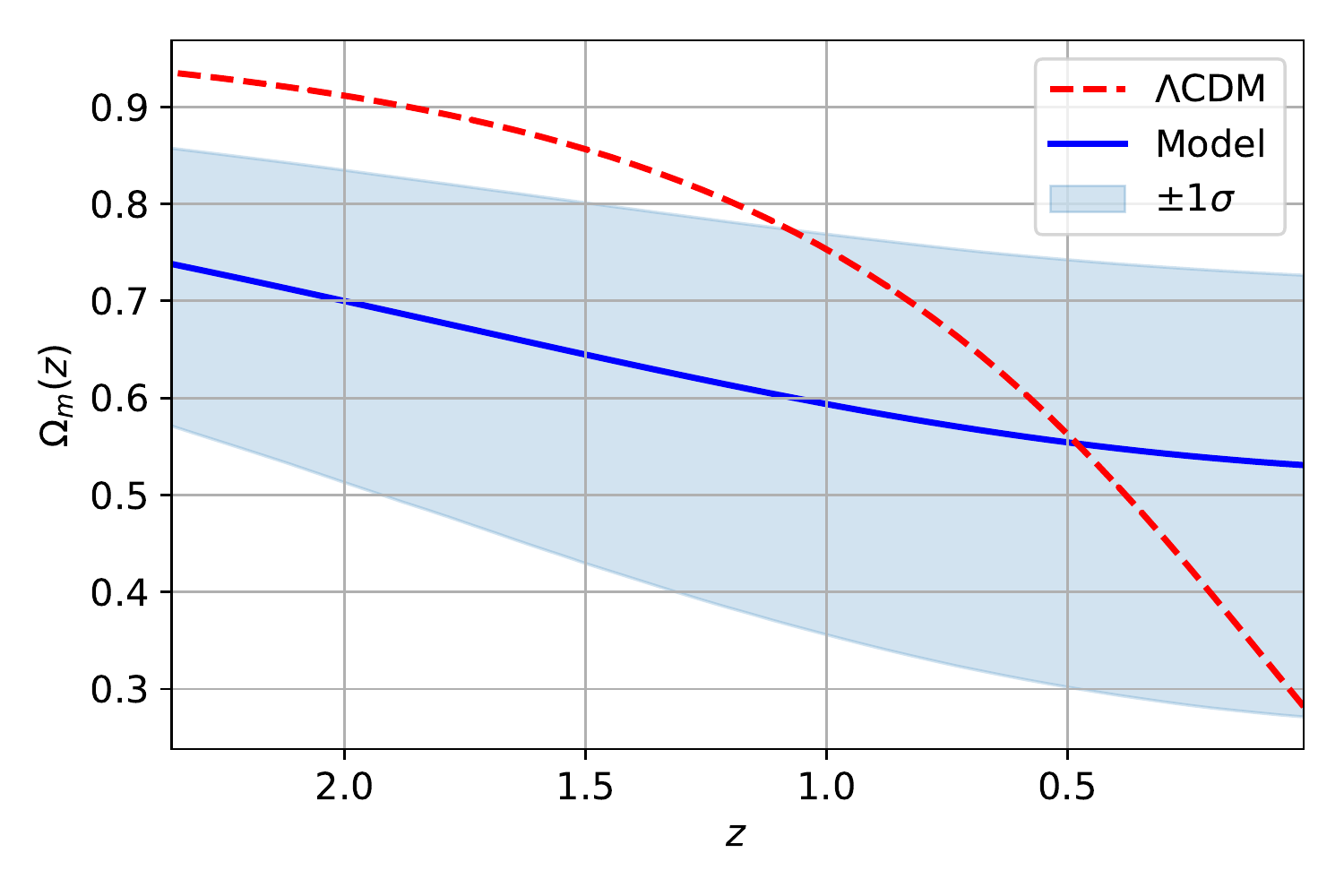}
    \caption{Matter density parameters for the $\Lambda$CDM model (red dashed line) and the fractional cosmological model (solid blue line) as a function of the redshift $z$. The~shaded curve represents the confidence region of the matter density parameter for the fractional cosmological model at a $1\sigma(68.3\%)$  confidence level (CL). The~figure was obtained using the best-fit values for the SNe Ia+OHD data presented in Table~\ref{tab:bestfits}.}
    \label{fig:OmFractional}
\end{figure}

\begin{figure}[H]
 %   \centering
    \includegraphics[scale = 0.65]{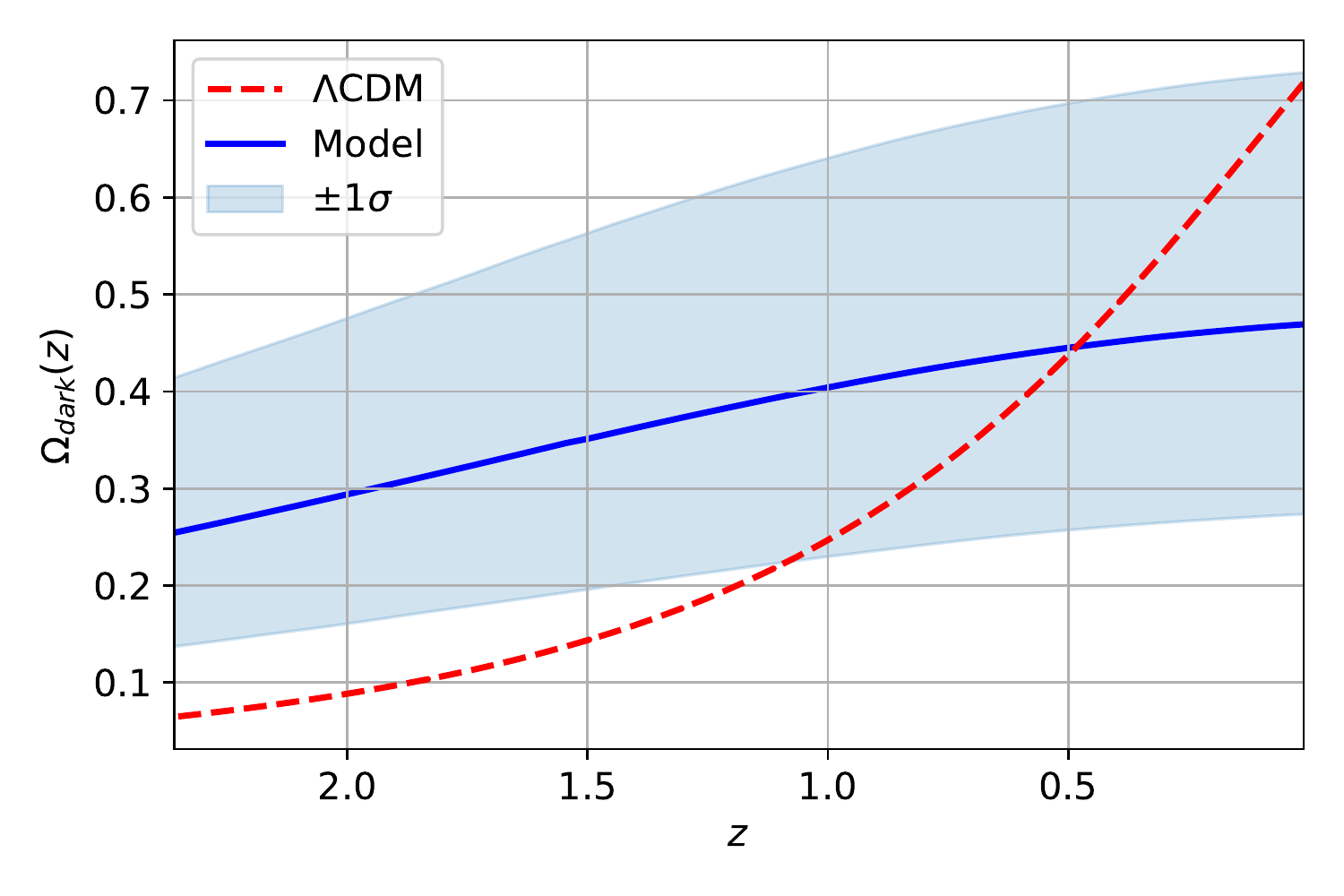}
    \caption{Dark energy density parameters for the $\Lambda$CDM model (red dashed line) and the fractional cosmological model (solid blue line) as a function of the redshift $z$. The~shaded curve represents the confidence region of the dark energy density parameter for the fractional cosmological model at a $1\sigma(68.3\%)$  confidence level (CL). The~figure was obtained using the best-fit values for the SNe Ia+OHD data presented in Table~\ref{tab:bestfits}.}
    \label{fig:OdFractional}
\end{figure}

Following this line, in~Figure~\ref{fig:qFractional}, we depict the deceleration parameter for the fractional cosmological model as a function of the redshift $z$ obtained from the best-fit values for the SNe Ia+OHD data presented in Table \ref{tab:bestfits}, with~an error band at $3\sigma$ CL. We also depict the deceleration parameter for the $\Lambda$CDM model as a reference model. From~this figure, we can conclude that the fractional cosmological model effectively experiences this transition at $z_{t}\gtrapprox 1$, with~the characteristic that $z_{t}>z_{t,\Lambda CDM}$, where $z_{t,\Lambda CDM}$ is the transition redshift of the $\Lambda$CDM model. Furthermore, the~current deceleration parameter of the fractional cosmological model is $q_{0}=-0.37_{-0.11}^{+0.08}$ at $3\sigma$ CL. On~the other hand, in~Figures~\ref{fig:OmFractional} and \ref{fig:OdFractional}, we depict the matter density and fractional density parameters for the fractional cosmological model (the last one interpreted as dark energy), respectively, as~a function of the redshift $z$ for~the best-fit values for the SNe Ia+OHD data presented in Table \ref{tab:bestfits}, with~an error band at $1\sigma$ CL. We depict the matter density and dark energy density parameters in both figures for the $\Lambda$CDM model. From~Figure~\ref{fig:OmFractional}, we can see that the matter density parameter for the fractional cosmological model, obtained from Equation \eqref{(71)}, presents significant uncertainties, which could be a consequence of their reconstruction from a Hubble parameter that does not take into account any EoS. In~this sense, the~current value of this matter density parameter at $1\sigma$ CL is $\Omega_{m,0}=0.531_{-0.260}^{+0.195}$, a~value that is in agreement with the asymptotic value obtained from Equation \eqref{Ominfty} of $\Omega_{m,t\to\infty}=0.519_{-0.262}^{+0.199}$, computed at $1\sigma$ CL for the best-fit values for the SNe Ia+OHD data presented in Table \ref{tab:bestfits}. Therefore, this larger value of $\Omega_{m,0}$ for the fractional cosmological model can, in~principle, explain the lower value of the current deceleration parameter $q_{0}$ and the excess of matter in the effective term $\rho_{\text{frac}}=3(\mu-1)t^{-1}H$ with $\Omega_{\text{frac}}(\alpha(s))=(\mu-1)/\alpha(s)$ (see Section~\ref{Sect:III.2}). Note that the current value $\Omega_{\text{frac},0}$ can be interpreted as the dark energy density parameter for the fractional cosmological model as $\Omega_{\text{frac},0}=0.469_{-0.195}^{+0.260}$, which satisfies the condition $\Omega_{m,0}+\Omega_{\text{frac},0}=1$.

\vspace{-6pt}

\unskip

Finally, we compute the cosmographic parameter known as the jerk, which quantifies if the fractional cosmological model tends to $\Lambda$ or if it another kind of DE, which can be written as
\begin{align}
    &j(s)=q(s)(2q(s)+1)-\frac{dq(s)}{ds}, 
\end{align}
where $q$ is given by Equation \eqref{DecelerationFinal}. Hence,
\begin{align}
    &j(\alpha(s))= \frac{12 (\mu -4)}{\alpha (s)}+\frac{(\mu -21) \mu +50}{\alpha
   (s)^2}-\frac{2 (\mu -3) (\mu -2) (\mu -1)}{\alpha (s)^3}+10. \label{Jerk}
\end{align}

\begin{figure}[H]
  %  \centering
    \includegraphics[scale = 0.63]{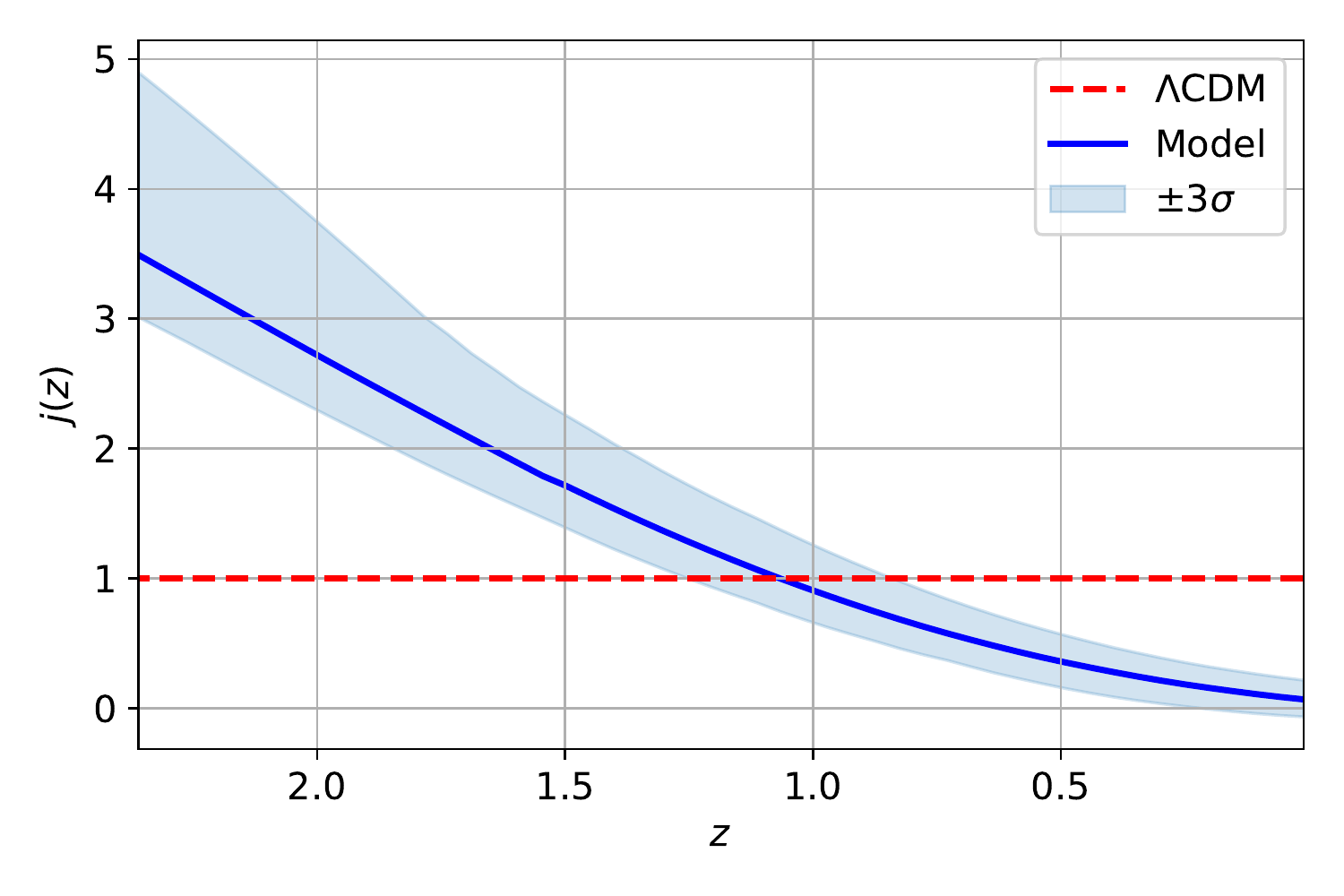}
    \caption{Jerks for the $\Lambda$CDM model (red dashed line) and the fractional cosmological model (solid blue line) as a function of the redshift $z$. The~shaded curve represents the confidence region of the jerk for the fractional cosmological model at a $3\sigma(99.7\%)$  confidence level (CL). The~figure was obtained using the best-fit values for the SNe Ia+OHD data presented in Table~\ref{tab:bestfits}.}
    \label{fig:JerkFractional}
\end{figure}

Figure~\ref{fig:JerkFractional}  represents the jerks for the $\Lambda$CDM model (red dashed line) and the fractional cosmological model (solid blue line) as a function of the redshift $z$. The~figure was obtained using the best-fit values for the SNe Ia+OHD data presented in Table~\ref{tab:bestfits} with an error band at $3\sigma$ CL, represented by a shaded region. A~departure of more than a $3\sigma$  CL for the current value for $\Lambda$CDM shows an alternative cosmology with an effective dynamical equation of state for the Universe for late times in contrast to $\Lambda$CDM.

\begin{figure}[H]
 %   \centering
    \includegraphics[scale = 0.63]{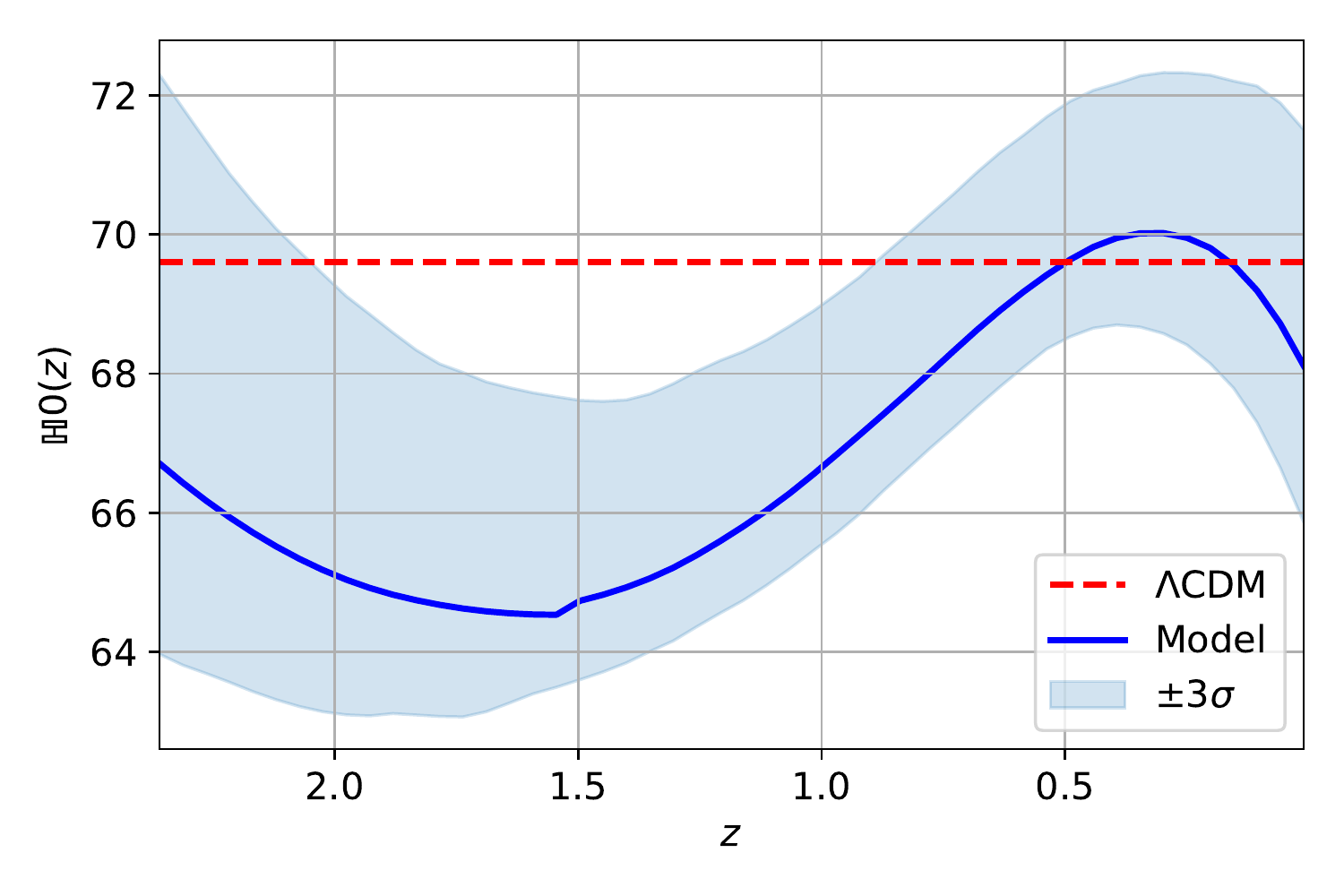}
    \caption{$\mathbf{\mathbb{H}}0$ diagnostics for the $\Lambda$CDM model (red dashed line) and the fractional cosmological model (solid blue line) as a function of the redshift $z$. The~shaded curve represents the confidence region of the $\mathbf{\mathbb{H}}0$ diagnostic for the fractional cosmological model at a $3\sigma(99.7\%)$  confidence level (CL). The~figure was obtained using the best-fit values for the SNe Ia+OHD data presented in Table~\ref{tab:bestfits}.}
    \label{fig:H0diagnostic}
\end{figure}

On the other hand, for~the reconstruction of the $\mathbf{\mathbb{H}}0(z)$ diagnostic \citep{H0diagnostic:2021} for the fractional cosmology, we define
\begin{align}
    \mathbf{\mathbb{H}}0(z)= \frac{H(z)}{\sqrt{\Omega_{m,0}(1+z)^{3}+1-\Omega_{m,0}}}, \label{HH00diagnostic}
\end{align}
where the Hubble parameter is obtained numerically by $H(z)=H_{th}(z)$ as we explained before. Therefore, in~Figure~\ref{fig:H0diagnostic}, we depict $\mathbf{\mathbb{H}}0$ diagnostic for the $\Lambda$CDM model (red dashed line) and the fractional cosmological model (solid blue line) as a function of the redshift $z$. The~figure was obtained using the best-fit values for the SNe Ia+OHD data presented in Table~\ref{tab:bestfits}, with~an error band at $3\sigma$ CL, represented by a shaded region. As~a reminder, in~both Figures~\ref{fig:JerkFractional} and \ref{fig:H0diagnostic}, we also depict the jerk and the $\mathbf{\mathbb{H}}0$ diagnostic for the $\Lambda$CDM model as a reference~model.

\unskip

\section{Conclusions}
\label{Sect:IX}
In this paper, we investigated the cosmological applications of power-law solutions of the type $a=(t/t_0)^{\alpha_0}$ in fractional cosmology, where $\alpha_0=t_0 H_0$ is the current age parameter. Additionally, given $\mu$,  the~order of the fractional derivative, and~$w$, the~matter equation of state, we have imposed compatibility conditions which  allow particular solutions to  $(\mu, w)$. That means that any solution of power law type is indeed a particular exact solution of the system, e.g.,~solution \eqref{3} for $\rho=0$; 
solution \eqref{17} for dust matter where $\alpha$ is defined by \eqref{zeta}; solutions $H_{1,2}$ defined by \eqref{solH-1}, with  $\alpha_\pm$ defined by \eqref{apm-1} and~where the compatibility conditions  \eqref{compatibility_1}
and \eqref{compatibility_2} are satisfied simultaneously for $\mu$ and $w$; and~the solution \eqref{exactRicatti22} such that $\mu$ and $w$ satisfy \eqref{compatibility_21} and \eqref{compatibility_22}, respectively. However, they are not the general solution. Moreover, we are interested in an exact solution that gives the general solution of the system. For~this purpose,  we solved the Riccati equation~\eqref{Riccati} independent of the EoS, where the solution for the scale factor is a combination of power laws,  i.e., \eqref{V}.  This solution is analyzed in Section~\ref{Gen_Sol_H}. 

Finally, combining the solution of Bernoulli’s Equation~\eqref{eqm} and the~inequality \eqref{ineq}, solving the differential inequality \eqref{eqN} and~approximating the different quadrature, we have obtained the approximate analytical solution  $E_{\text{approx}}(z)$ given by \eqref{Ez}, and~$t_{\text{approx}}(z)$ given by \eqref{tz} as $z\rightarrow -1$, where $\alpha_0$ is defined by \eqref{alpha_0}, 
$A$ is defined by \eqref{(EQ140)} and~satisfies \eqref{inequality} and~$m_0= \alpha_0^2-A^2$. This is an accurate approximation of $E(z)$ as $z\rightarrow -1$ provided  $2\leq \mu \leq \frac{1}{10} \left(63+\sqrt{849}\right)\lesssim 9.21376$.

Finally, we estimated the free parameters $(\alpha_0, \mu)$ using cosmological data and the re-parameterization $H_{0}=100\frac{\text{km/s}}{\text{Mpc}}h$,  $\alpha_0=  \frac{1}{6} \left(9 -2 \mu +\sqrt{8 \mu  (2 \mu -9)+105}\right)(1+ 2 \epsilon_0)$. 

Separate analyses of the SNe Ia data and OHD, and the joint analysis with  SNe Ia data + OHD, led, respectively, to $h=0.696_{-0.295}^{+0.302}$, $\mu=1.340_{-0.339}^{+2.651}$ and  $\epsilon_0=\left(1.976_{-2.067}^{+1.709}\right)\times 10^{-2}$;   $h=0.675_{-0.021}^{+0.041}$, $\mu=2.239_{-1.190}^{+1.386}$ and $\epsilon_0=\left(0.865_{-0.773}^{+0.793}\right)\times 10^{-2}$; and~ $h=0.684_{-0.027}^{+0.031}$, $\mu=1.840_{-0.773}^{+1.446}$ and $\epsilon_0=\left(1.213_{-1.057}^{+0.482}\right)\times 10^{-2}$, where the best-fit values were calculated at $3\sigma$ CL. On~the other hand, these best-fit values led to an age of the Universe with a value of $t_0=\alpha_0/H_0=25.62_{-4.46}^{+6.89}\;\text{Gyrs}$, a~current deceleration parameter of $q_{0}=-0.37_{-0.11}^{+0.08}$ (both at $3\sigma$ CL) and~a current matter density parameter of $\Omega_{m,0}=0.531_{-0.260}^{+0.195}$ at $1\sigma$ CL. Finding a Universe roughly twice as old as the one of $\Lambda$CDM is a distinction of fractional cosmology. Focusing our analysis on these results, we can conclude that the region in which $\mu>2$ is not ruled out by observations. This region of a parameter is relevant because, in~the absence of matter, fractional cosmology gives a power-law solution $a(t)= \left(t/t_0\right)^{\mu-1}$, which is accelerated for $\mu>2$. We presented a fractional origin model that leads to an accelerated state without appealing to $\Lambda$ or dark~energy. 

%%%%%%%%%%%%%%%%%%%%%%%%%%%%%%%%%%%%%%%%%%
%% Optional
%\appendixtitles{yes} % Leave argument "no" if all appendix headings stay EMPTY (then no dot is printed after "Appendix A"). If the appendix contains a heading, change the argument to "yes".
%\appendixstart
%\appendix
%\section[\appendixname~\thesection]

\vspace{6pt}

%%%%%%%%%%%%%%%%%%%%%%%%%%%%%%%%%%%%%%%%%%
%% optional
%\supplementary{The following supporting information can be downloaded at:  \linksupplementary{s1}, Figure S1: title; Table S1: title; Video S1: title.}


%%%%%%%%%%%%%%%%%%%%%%%%%%%%%%%%%%%%%%%%%%
\section*{Author contributions}
Conceptualization, E.G. and G. L.; methodology, E.G., G.L. and G.F.A.; software,  E.G. and G.L.; validation, E.G., G.L. and G.F.A.; formal analysis, E.G., G.L. and G.F.A.; investigation, E.G., G.L. and G.F.A.; resources, G.L.;  writing---original draft preparation, G.L.; writing---review and editing, E.G., G.L. and G.F.A.; visualization,  E.G. and G.L.; supervision, G.L.; project administration, G.L.; funding acquisition, G.L.. All authors have read and agreed to the published version of the manuscript.

\section*{Funding}

G. L. was funded by Vicerrectoría de Investigación y Desarrollo Tecnológico (VRIDT) at Universidad Católica del Norte through Concurso De Pasantías De Investigación Año 2022, Resolución VRIDT No. 040/2022 and Resolución VRIDT No. 054/2022.

\section*{Data availability}
The data supporting this article can be found in Section~\ref{Sect:VII}.

\acknowledgments{The authors are thankful for the support of Núcleo de Investigación Geometría Diferencial y Aplicaciones, Resolución VRIDT No. 096/2022.  The~authors thank Samuel Lepe for initial discussions. E.G. acknowledges the support of Dirección de Investigaci\'on y Postgrado at Universidad de~Aconcagua. GFA acknowledges support from DINVP and Universidad Iberoamericana.}

\section*{Conflicts of interest}
The authors declare no conflicts of interest. The~funders had no role in the design of the study; in the collection, analyses, or~interpretation of data; in the writing of the manuscript; or in the decision to publish the~results.

\end{document}